# Effect of Ionic Advection on Electroosmosis over Charge Modulated Surfaces: beyond the Weak Field Limit

Uddipta Ghosh and Suman Chakraorty[1]

[1]Department of Mechanical Engineering, Indian Institute of Technology Kharagpur,
Kharagpur, West Bengal, India – 721 302

**ABSTRACT**

The present study deals with the effect of ionic advection on electroosmotic flow over charge modulated surfaces in a generalized paradigm when the classically restrictive "weak field" limit may be relaxed. Going beyond the commonly portrayed weak field limit (i.e, the externally applied electric field is overweighed by the surface-induced electrical potential, towards charge distribution in an electrified wall-adhering layer) for electroosmotic transport, we numerically solve the coupled full set of Poisson-Nernst-Planck (PNP) – Navier Stokes equations, in a semi-infinite domain, bounded at the bottom by a charged wall. Further, in an effort to obtain deeper physical insight, we solve the simplified forms of the relevant governing equations for low surface potential in two separate asymptotic limits: (i) a regular perturbation solution for Low Ionic Peclet number ($Pe$), where $Pe$ is employed as the gauge function and (ii) a matched asymptotic solution for $O(1)$ Pe in the Thin Electric Double Layer (EDL) limit. We demonstrate that reasonably good agreement is observed between the analytical and numerical solutions. Our analysis reveals that the primary effect of $Pe$ on the flow is to slow down the "free stream velocity", adding to the periodicity of the flow, while it also induces significant changes in the overall potential. We further show that the electrical double layer thickness strongly dictates the "free stream velocity", and a simple Smoluchowski type of slip boundary condition cannot be used, if the effect of advection is taken into account. These results can be of significant importance in designing microfluidic and nanofluidic systems with surface charge modulation.

## 1. Introduction

Electroosmosis (Hunter 2013), defined as the motion of a fluid relative to an electrically charged substrate under the action of an externally applied electric field, has long been an area of interest, primarily for its wide spectrum of applications, starting from efficient chip cooling (Stone et al. 2004) to bringing about efficient mixing in narrow confinements (Sugioka 2010). Electroosmosis has also been widely popular in lab-on-a-hip based microfluidic devices (see for example the references, (Mark et al. 2010; Haeberle & Zengerle 2007; Srinivasan et al. 2004; Chin et al. 2007) and the references therein), since it removes the necessity of having complex moving devices as a part of the flow actuating mechanism, as classically required for pressure driven flows. The strength of flow in electroosmosis strongly depends on the magnitude of the potential at the solid-liquid interface, commonly known as the "zeta potential" (Hunter 2013). More recently, a number of researchers have demonstrated (Ajdari 1995; Bahga et al. 2010; Ghosh & Chakraborty 2012; Mortensen et al. 2005; Sugioka 2010; Zhang et al. 2006) that patterning the surface potential or charge can lead to a wide variety of flow dynamics, which can be used for efficient mixing in microchannels (Ghosh & Chakraborty 2012; Zhang et al. 2006; Erickson & Li 2002; Chen & Cho 2007; Chang & Yang 2008).

A number of studies (Ajdari 1996; Ajdari 1995; Ajdari 2001; Bahga et al. 2010; Chang & Yang 2008; Chen & Cho 2007; Ghosal 2002; Ghosh & Chakraborty 2013; Ghosh & Chakraborty 2012; Mortensen et al. 2005; Naga Neehar & Chakraborty 2011; Ng & Chu 2011; Stroock et al. 2002; Sugioka 2010; Zhang et al. 2006; Erickson & Li 2002; Zhao 2011) have focused on various aspects of electroosmotic flows in presence of patterned surface potential. These include evaluation of electroosmotic mobility tensor over a stick-slip surface (Ng & Chu 2011; Bahga et al. 2010), effects of modulated slip and potential on mixing in narrow confinements (Ghosh & Chakraborty 2012; Chang & Yang 2008; Chen & Cho 2007; Erickson & Li 2002; Zhang et al. 2006), to name a few. However, almost all of these studies start from the Poisson-Boltzmann equation to describe the charge distribution in the channel and the resulting electroosmotic body force on the fluid. A more fundamental approach is obviously to start from the Nernst-Planck equations, which take into account the advection of ions, along with diffusion and electromigration. The strength of advection of ions is indicated by the ionic Péclet number ($Pe$). If $Pe$ is assumed to be small enough, one recovers the classical Poisson-Boltzmann equation, thus breaking the complex two-way coupling between electromechanics and fluid flow.

Although the Reynolds number ($Re$) remains typically small for electroosmotic flows in narrow confinements, the ionic Péclet number, $Pe = Re.Sc$ ($Sc$ being Schmidt number, defined as the ratio of the kinematic viscosity to the ionic diffusivity) can still be O(1) or larger, since $Sc$ usually has a large value for typical ionic species. This effect, however, does not create any alteration in the resultant flow physics in case of uniform surface charge, primarily attributable to a fortunate orthogonality between the directions of surface electrification (and hence the direction of charged species concentration gradient) and fluid flow. However, in case of charge modulated surfaces, the flow-field is inherently multi-dimensional. Therefore, it can be inferred that, despite the fluid flow being viscosity dominated, the advection of ions in the Nernst-Planck equation can still be pivotal in determining the electro-hydrodynamic transport. This can be attributed to on-stream as well as cross-stream advection of the charged species in case of charge modulated surfaces. Accordingly, two-way coupling of ionic transport and fluid flow may take place, which does not otherwise feature in the classical scenarios of electroosmotic flow with uniform surface potential. In addition, since in presence of charge modulated surfaces, the orthogonality between the flow direction and the direction of surface potential distribution gets destroyed, the externally applied axial field may significantly alter the charge distribution inside the electrical double layer (EDL). Accordingly, in addition to accounting for the charge advection effect, the commonly used "weak field limit" may need to be relaxed for electroosmosis over charge modulated surfaces.

Recently, Zhao and co-workers (Zhao 2010; Zhao & Bau 2008) have investigated electroosmotic mobility (Zhao 2010) over a stick-slip surface with patterned potential, in presence of ionic advection. However, in their works, only a weak field limit was considered. As per this consideration, the effect of the externally applied electric field may be neglected towards determining the potential distribution within the EDL, the latter being fully attributed to the substrate-induced electrostatic potential. Following this consideration, regular perturbation equations were derived taking the external field strength as the gauge function. Moreover, their work focused solely on the effects of stick slip surface on the electroosmotic mobility. On the other hand, electrokinetics beyond weak field limit has been extensively addressed only recently (Schnitzer & Yariv 2014; Schnitzer & Yariv 2012; Yariv 2005), albeit for the cases of electrophoresis of small particles and Induced charge electroosmosis with uniform surface potential, under the thin EDL limit. A close review of the concerned literature thus reveals that the effect of ionic advection on electroosmosis, in presence of patterned surface potential, within and beyond the limits of thin EDL and weak axial field is yet to be properly addressed. As already mentioned, such a physical paradigm makes the classical Poisson-Boltzmann distribution invalid and can potentially lead to interesting flow patterns, through the inherent complicated interplay between hydrodynamics and ionic transport. It may be noted at this point that the relaxation of weak field limit complicates the analysis of electroosmosis with ionic advection to a significant extent, because of an associated complex interplay between the applied electric field, induced electric field, ionic concentration distributions and fluid flow. Such complicated inter-connections are significantly simplified in case charge advection is considered within the paradigm of "weak field limit".

With the aforementioned motivation in mind, here we attempt to analyze the flow dynamics along with the charge and potential distribution for electroosmotic flows in a semi-infinite electrolyte region over charge modulated surfaces for arbitrary EDL thickness, beyond the weak field limit, while also accounting for ionic advection. The analysis can be largely divided into two parts: first the full system of relevant governing equations is numerically solved for low to moderate values of the surface potential, for a wide range of EDL thickness, ionic Péclet number and axial field strength. Next, we attempt to solve the governing equations, for two separate asymptotic limits. First, we derive regular asymptotic solutions, using *Pe* as the gauge function for low surface potential and low values of *Pe* (weak advection), assuming arbitrary EDL thickness. Second, we employ a matched asymptotic expansion technique in the limit of low surface potential and thin EDL, assuming *Pe* to be O(1) in the process. In both the limits we relax the paradigm of "weak field" approximation. We compare the analytical and numerical results within appropriate parametric range and demonstrate that they show reasonably good agreement. The central finding of our analysis is that, effects of field strength and advection strength have opposing effects on the development of the overall flow field. We demonstrate that the presence of a strong axial field accelerates the "free stream" or "far field" velocity, whereas inclusion of ionic advection decreases the same. Our matched asymptotic solutions specifically reveal that the dominant effect of interaction between advection and EDL is to slow down the "free-stream" velocity, which is in perfect agreement with the numerical predictions. We further depict that the effect of advection on the free stream velocity and over all flow dynamics is most pronounced at certain EDL thicknesses, while they are almost negligible for thick EDLs, even for relatively moderate values of *Pe*.

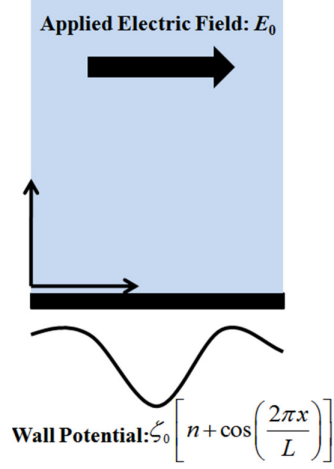

Figure 1: A schematic representation of system geometry. We consider a semi-infinite electrolyte solution lying on top of a solid surface. The origin, the *x* and the *y* axes have been depicted in the figure. An electric field of magnitude $E_0$ is applied in the *x* direction to actuate the flow. The surface bears a potential of the form: $\zeta = \zeta_0 \left( n + \cos\left( \frac{2\pi x}{L} \right) \right)$

## 2. System Description and Governing Equations

### *2.1 System Description*

We consider a symmetric (*z:z*) semi-infinite electrolyte solution in contact with a solid surface, as shown in the schematic figure 1. The surface bears a potential of the form, $\zeta = \zeta_0 \left( n + \cos\left( \frac{2\pi x}{L} \right) \right)$ while an electric field of magnitude $E_0$ is applied in the *x* direction to actuate the flow. Since the fluid contains electrolytic solution, an EDL is formed near the solid surface, which imparts a net charge to the fluid. We denote the characteristic EDL thickness by $\lambda_D$, which has the following expression: $\lambda_D = \sqrt{\frac{\varepsilon kT}{2c_0 z^2 e^2}}$, where $\varepsilon$ is the permittivity of the fluid medium and $c_0$ is the bulk electrolyte concentration, while the rest of the symbols bear their usual meanings. In general, thin EDL limit consideration would implicate $\delta = \lambda_D / L \ll 1$, and the classical weak field approximation would indicate, $\beta = 2\zeta_0\pi/E_0 L \gg 1$. However, in the present study, in order to investigate the flow patterns for a more general system, we assume both $\beta$ and $\delta$ to have arbitrary values, relaxing the above constraints.

Before we move to the governing equations, it is important here to state the assumptions undertaken in the present analysis. Firstly, we assume the inertial effects on the fluid flow to be negligible as compared to the viscous forces, i.e., $Re \ll 1$, an assumption, which has been widely employed in the literature concerning electroosmosis (Ajdari 1995; Ajdari 1996; Ajdari 2001; Bahga et al. 2010). However, we do account for a non-vanishing ionic Pélet number. In an effort to validate such a consideration, we perform a simple order of magnitude analysis for the related parameters. For the present case, the Reynolds number is defined as, $Re = \dfrac{\rho u_c L}{\mu}$ where, $u_c$ is the characteristic velocity and $\rho$ is the fluid density. Taking water as the fluid, we find, $\rho \sim 1000$ kg/m$^3$, $\mu = 10^{-3}$ P, while typical electroosmotic velocities lie in the range of $u_c \sim 0.1 - 1$ mm/s. Therefore, for a modulation wavelength $L \sim 1 - 10$ μm, one obtains, $Re \sim O(10^{-4}) - O(10^{-2})$, which shows that the fluid inertia can indeed be neglected. However, the ionic Péclet number is defined as, $Pe = \dfrac{u_c L}{D}$, where $D$ is the ionic diffusivity. Typical value of $D$ for electrolytes is usually in the tune of (Bazant et al. 2004; Højgaard Olesen et al. 2010) $D \sim 10^{-8} - 10^{-9}$ m$^2$/s. Therefore, for the aforementioned values of $u_c$ and $L$, one obtains, $Pe \sim O(10^{-2}) - O(10)$ or larger, depending on the values of zeta potential. Therefore, we conclude that, despite Re being small, the ionic Péclet number can still assume relatively large values, owing to very low ionic diffusivities. Secondly, we assume that the velocities, pressure, charge distribution, and the potential resulting from the surface charge are periodic along the axial direction. This assumption has also been used widely in a number of previous studies (Ajdari 1995; Bahga et al. 2010), while investigating electroosmosis in presence of periodic patterning.

## 2.2 Governing equations

The conservation equations for the ionic species are given by the Nernst-Planck equations, which can be expressed in the following form:

$$\mathbf{v} \cdot \nabla c_i = D \nabla^2 c_i \pm \dfrac{zeD}{kT} \nabla \cdot \{c_i \nabla \psi\} \qquad (2.1)$$

In (2.1), $\mathbf{v}$ is the fluid velocity, $i = 1$ denotes the positive ions, while $i = 2$ denotes the negative ions, whereas, $\psi$ is the total electrostatic potential. Additionally, in (2.1), the "+" sign is valid for $i = 1$, while the "-" sign remains valid for $i = 2$. The electrostatic potential additionally has to satisfy the Poisson equation, which reads:

$$-\nabla^2 \psi = \dfrac{ze(c_1 - c_2)}{\varepsilon} \qquad (2.2)$$

The electrostatic potential can be represented in the following form (Bahga et al. 2010):

$$\psi = -E_0 x + \varphi \qquad (2.3)$$

In equation (2.3), the $-E_0 x$ is the contribution from the externally applied potential, whereas, $\varphi$ is the potential induced due to the charges present on the surface. In other words $\varphi$ is the EDL potential. Enforcing (2.3) into (2.1) and (2.2), one obtains:

$$\mathbf{v} \cdot \nabla c_i = D \nabla^2 c_i \pm \dfrac{zeD}{kT} \left[ \nabla \cdot \{c_i \nabla \varphi\} - \dfrac{\partial c_i}{\partial x} \mathbf{e_x} \right] \qquad (2.4)$$

$$-\nabla^2 \varphi = \dfrac{ze(c_1 - c_2)}{\varepsilon} \qquad (2.5)$$

Equations (2.4) and (2.5) are subject to the following boundary conditions: At the wall, the ionic species concentrations satisfy no flux conditions into the surface, while the potential $\varphi$ is specified:

At $y = 0$, $\dfrac{\partial c_i}{\partial y} \pm \dfrac{ze}{kT} c_i \dfrac{\partial \varphi}{\partial y} = 0$ (2.6)

At $y = 0$, $\varphi = \zeta_0 \left( n + \cos\left( \dfrac{2\pi x}{L} \right) \right)$ (2.7)

Far away from the wall, both charge density and the EDL potential would vanish, which reads:

$\varphi \to 0, c_i \to c_0 \, (i = 1, 2)$, as $y \to \infty$ (2.8)

Along the axial direction, all the variables are assumed to satisfy periodic condition. The fluid velocities satisfy the Stokes equation (considering an electroosmotic body force) and the continuity equation, which have the following form:

$0 = -\nabla p + \mu \nabla^2 \mathbf{v} + \varepsilon \nabla^2 \varphi (\nabla \varphi - E_0 \mathbf{e_x}); \quad \nabla \cdot \mathbf{v} = 0$ (2.9)

Equations (2.9) are subject to the following boundary conditions:

At $y = 0$, $\mathbf{v} = 0$ – no slip and no penetration (2.10)

At $y \to \infty$, $v \to 0$, $\dfrac{\partial u}{\partial y} \to 0$ (2.11)

We now non-dimensionalize the governing equations, in the following way: for any variable $\xi$ we write, $\bar{\xi} = \xi / \xi_c$, where $\xi_c$ is the reference/characteristic value for the variable under consideration. We chose the following char. values: $x_c = y_c = d = L/2\pi$, $c_{i,c} = c_0$, $\varphi_c = \zeta_0$, $u_c = v_c = u_{HS}$, $p_c = \mu u_c/d$, where, $u_{HS}$ is the Smoluchowski velocity given by $u_{HS} = \dfrac{\varepsilon \zeta_0 E_0}{\mu}$. We additionally, define the following non-dimensional ratios and numbers: $\delta = \dfrac{\lambda_D}{d}, \beta = \dfrac{\zeta_0}{E_0 d}, \bar{\zeta}_0 = \dfrac{ze\zeta_0}{kT}$ and Péclet number: $Pe = \dfrac{u_c d}{D}$. For ease of representation, we recast the Nernst-Planck equations with a new set of dependent variables (Bazant et al. 2004; Højgaard Olesen et al. 2010):

Net charge density: $\bar{\rho}_e = \bar{c}_1 - \bar{c}_2$ and total electrolyte concentration: $\bar{c} = \bar{c}_1 + \bar{c}_2$ (2.12).

Enforcing the above non-dimensionalization scheme and new variables defined in (2.12), in (2.4 – 2.5) and (2.9), one can write:

$Pe \left( \bar{\mathbf{v}} \cdot \bar{\nabla} \bar{c} \right) = \bar{\nabla}^2 \bar{c} + \bar{\zeta}_0 \bar{\nabla} \cdot \left( \bar{\rho}_e \bar{\nabla} \bar{\varphi} \right) - \dfrac{\bar{\zeta}_0}{\beta} \dfrac{\partial \bar{\rho}_e}{\partial \bar{x}}$ (2.13)

$$Pe\left(\bar{\mathbf{v}} \cdot \bar{\nabla}\bar{\rho}_e\right) = \bar{\nabla}^2 \bar{\rho}_e + \bar{\zeta}_0 \bar{\nabla} \cdot \left(\bar{c}\bar{\nabla}\bar{\varphi}\right) - \frac{\bar{\zeta}_0}{\beta}\frac{\partial \bar{c}}{\partial \bar{x}} \tag{2.14}$$

$$\bar{\nabla}^2 \bar{\varphi} = -\frac{\bar{\rho}_e}{2\bar{\zeta}_0 \delta^2} \tag{2.15}$$

$$-\bar{\nabla}\bar{p} + \nabla^2 \bar{\mathbf{v}} + \bar{\nabla}^2 \bar{\varphi}\left(\beta\bar{\nabla}\bar{\varphi} - \mathbf{e}_\mathbf{x}\right) = 0 \tag{2.16}$$

$$\bar{\nabla} \cdot \bar{\mathbf{v}} = 0 \tag{2.17}$$

The non-dimensional boundary conditions take the form:

$$\frac{\partial \bar{c}}{\partial \bar{y}} + \bar{\zeta}_0 \bar{\rho}_e \frac{\partial \bar{\varphi}}{\partial \bar{y}} = 0, \text{ at } \bar{y} = 0 \tag{2.18}$$

$$\frac{\partial \bar{\rho}_e}{\partial \bar{y}} + \bar{\zeta}_0 \bar{c} \frac{\partial \bar{\varphi}}{\partial \bar{y}} = 0, \text{ at } \bar{y} = 0 \tag{2.19}$$

$$\bar{\varphi} = n + \cos \bar{x}, \text{ at } \bar{y} = 0 \tag{2.20}$$

$$\mathbf{v} = 0, \text{ at } \bar{y} = 0 \tag{2.21}$$

$$\text{at, } \bar{y} \to \infty, \bar{\phi} \to 0, \bar{\rho}_e \to 0, \bar{c} \to 2, \bar{v} \to 0 \text{ and } \frac{\partial \bar{u}}{\partial \bar{y}} \to 0 \tag{2.22}$$

Along *x*, all the above variables are assumed to satisfy periodic boundary condition. Equations (2.13) – (2.17), subject to the conditions (2.18 – 2.22), represent the complete set of governing equations describing the electrokinetics and the fluid motion. Note that in (2.13) and (2.14), $\beta \gg 1$ represents the weak field limit, wherein, the last terms in the equations under consideration vanish and one recovers the equations pertaining to "weak field approximation".

## 3. Linearized Poisson-Nernst-Planck (PNP) equations for low surface potential

The low potential limit occurs when the potential satisfies the criterion (Bahga et al. 2010): $|ze\varphi/kT| \sim |ze\zeta_0/kT| \ll 1$. This condition is usually met when, $|\zeta_0| \ll 25\,mV$. In such cases, the total electrolyte concentration only slightly deviates from the bulk value and the net charge density is approximately proportional to $\bar{\zeta}_0$. Therefore, we can significantly simplify the Nernst-Planck equations by linearizing the same, which enables us to obtain analytic solutions for various asymptotic limits. To this end, following Mortensen et al., (Mortensen et al. 2005) and Bazant et al., (Bazant et al. 2004) the individual species concentrations can be expressed as follows:

$$\bar{c}_1 = 1 + \delta\bar{c}_1; \bar{c}_2 = 1 - \delta\bar{c}_2; \text{ with } |\delta\bar{c}_i| \sim O\left(\bar{\zeta}_0\right) \ll 1, i = 1, 2 \tag{3.1}$$

Therefore, the charge density becomes (Mortensen et al. 2005), $\bar{\rho}_e = \delta\bar{c}_1 - \delta\bar{c}_2 \sim O\left(\bar{\zeta}_0\right)$ and the total electrolyte concentration becomes (Mortensen et al. 2005): $\bar{c} \approx 2 + O\left(\bar{\zeta}_0^2\right)$. Enforcing these simplifications in (2.13 – 2.15) and subsequently neglecting all the O($\bar{\zeta}_0^2$) or

higher order terms, the resulting linearized PNP equations can be written in the following way:

$$Pe\left(\bar{\mathbf{v}} \cdot \bar{\nabla}\bar{\rho}_e\right) = \bar{\nabla}^2 \bar{\rho}_e + 2\bar{\zeta}_0 \bar{\nabla}^2 \bar{\varphi} \tag{3.2}$$

$$\delta^2 \bar{\nabla}^2 \bar{\varphi} = -\frac{\bar{\rho}_e}{2\bar{\zeta}_0} \tag{3.3}$$

In (3.2) and (3.3), we have assumed $\beta \sim O(1)$. It is interesting to note that, in the low potential limit, the externally applied field has no direct effect on the charge distribution, even beyond the weak field limit, which is a natural consequence of neglecting $O(\bar{\zeta}_0^2)$ or higher order terms. The resulting fluid flow equations can be written in the following way:

$$-\bar{\nabla}\bar{p} + \bar{\nabla}^2 \bar{\mathbf{v}} + \bar{\nabla}^2 \bar{\varphi}\left(\beta\bar{\nabla}\bar{\varphi} - \mathbf{e}_\mathbf{x}\right) = 0 \tag{3.4}$$

$$\bar{\nabla} \cdot \bar{\mathbf{v}} = 0 \tag{3.5}$$

The boundary condition (2.18) is rendered useless, while the simplified form of (2.19) reads:

$$\frac{\partial \bar{\rho}_e}{\partial \bar{y}} + 2\bar{\zeta}_0 \frac{\partial \bar{\varphi}}{\partial \bar{y}} = 0 \tag{3.6}$$

Rest of the boundary conditions remain same as expressed in (2.20 – 2.22). It is interesting to note that, the low potential limit allows us to solve one less equation, since the equation for the total electrolyte concentration is not required on account of the assumption that $\bar{c} \approx 2$. We further note that the next correction to the concentration dependent on the potential is $O(\bar{\zeta}_0^2)$. Observing (3.2 – 3.5), we conclude that these linearized equations can be solved in two separate asymptotic limits. First, we can assume $\delta$ (EDL thickness) to be arbitrary (not more than $O(1)$), which enables us to employ a simple regular perturbation expansion for low Peclet numbers ($Pe \ll 1$), using $Pe$ as the gauge function. For such cases, the asymptotic expansion of any variable $\xi$, reads:

$$\bar{\xi} = \bar{\xi}_0 + Pe\bar{\xi}_1 + Pe^2\bar{\xi}_2 + O\left(Pe^3\right) \tag{3.7}$$

In (3.7), $\xi$ can represent variables like $u$, $v$, $p$, $\rho_e$ etc. A second and a more practically relevant and interesting choice is to assume $Pe$ to be arbitrary, not exceeding $O(1)$ in magnitude and $\delta \ll 1$. This leads to a thin EDL approximation, wherein all the charges and hence the actuating forces on the fluid is concentrated in a very thin region (EDL) near the wall. This region typically extends to a distance of $O(\delta)$ vertically above the wall. Therefore, in the limit $\delta \ll 1$, one has to apply a matched asymptotic expansion (singular perturbation) to resolve the dynamics inside the EDL. Then, there are two distinct layers: (i) the *Double layer*, or the *inner layer*, where the relevant length scale in the $y$ direction is $\delta$ and (ii) the *Bulk* or, the *outer layer*, where the relevant length scale remains $O(1)$ (or, $d$ in dimensional form). In each layer, any variable $\xi$ is expanded in $\delta$ as follows:

$$\bar{\xi} = \bar{\xi}_0 + \delta\bar{\xi}_1 + \delta^2\bar{\xi}_2 + O\left(\delta^3\right) \tag{3.8}$$

In (3.8), $\xi$ can represent variables like *u*, *v*, *p*, $\rho_e$ etc, in both EDL and bulk. The variables inside the EDL, satisfy the conditions at the wall, whereas, their counterparts in the bulk satisfy the far field conditions. For consistent solutions, one has to match the pertinent variables in the outer and inner layer, in their common region of validity. Note that, matched asymptotic method has previously been applied in a number of studies (Schnitzer & Yariv 2012; Yariv et al. 2011; Schnitzer et al. 2012; Ramos et al. 2003) to resolve the EDL, subject to a wide range of externally imposed conditions. In section 4, we deal with the first asymptotic limit, i.e., the low *Pe* limit (or, equivalently the *Weak Advection limit*) and in section 5, we address the more practically relevant and interesting thin EDL limit, which is perhaps more insightful as compared to the regular case of $Pe \ll 1$.

## 4. Asymptotic Limit – I: Low Ionic Peclet numbers ($Pe \ll 1$)

As noted earlier, in the limit $Pe \ll 1$, we will expand all the variables in equations (3.2 – 3.5) and the corresponding boundary conditions, using the asymptotic series given in (3.7). In the present section, we only derive a three term asymptotic series, i.e., we determine terms only upto $O(Pe^2)$.

### *4.1. Leading order solutions*

The leading order governing equations are as follows:

$$\bar{\nabla}^2 \bar{\rho}_e^{(0)} + 2\bar{\zeta}_0 \bar{\nabla}^2 \bar{\varphi}_0 = 0 \tag{4.1}$$

$$\bar{\nabla}^2 \bar{\varphi}_0 = -\frac{\bar{\rho}_e^{(0)}}{2\bar{\zeta}_0 \delta^2} \tag{4.2}$$

$$-\bar{\nabla}\bar{p}_0 + \nabla^2 \bar{\mathbf{v}}_0 + \bar{\nabla}^2 \bar{\varphi}_0 \left( \beta \bar{\nabla} \bar{\varphi}_0 - \mathbf{e_x} \right) = 0; \quad \bar{\nabla} \cdot \bar{\mathbf{v}}_0 = 0 \tag{4.3}$$

Subject to the boundary conditions:

$$\frac{\partial \bar{\rho}_e}{\partial \bar{y}} + 2\bar{\zeta}_0 \frac{\partial \varphi}{\partial \bar{y}} = 0, \quad \bar{\varphi}_0 = n + \cos(\bar{x}), \quad \mathbf{v}_0 = 0 \text{ at } \bar{y} = 0 \tag{4.4}$$

and at $\bar{y} \to \infty$, $\bar{\varphi}_0 \to 0, \bar{\rho}_e^{(0)} \to 0, \bar{v}_0 \to 0$ and $\frac{\partial \bar{u}_0}{\partial \bar{y}} \to 0 \tag{4.5}$

We first note that in the leading order, governing equations for potential and the charge density are decoupled from the momentum equations, and hence can be solved independently. Taking hint from the boundary conditions, one can easily reach at the following solutions for charge density and potential:

$$\bar{\rho}_e^{(0)} = -2\bar{\zeta}_0 \bar{\varphi}_0; \quad \bar{\varphi}_0 = n e^{-\bar{y}/\delta} + e^{-\bar{y}/\delta_1} \cos(\bar{x}) \tag{4.6}$$

In (4.6), $\delta_1 = \left[ \sqrt{1 + \delta^{-2}} \right]^{-1}$. We note that, the potential and the charge distribution in (4.6) is nothing but the Boltzmann distribution, subject to Debye-Huckel linearization. This is expected, since in the leading order, as evident from (4.1), charges are in equilibrium, which naturally results in Boltzmann distribution. The solutions in (4.6) can be plugged in (4.3) to evaluate the velocity fields, which can be expressed in the following form, with the help of a stream function: $\bar{u}_0 = \hat{u}_0(\bar{y}) + \partial S_0 / \partial \bar{y}$; $\bar{v}_0 = -\partial S_0 / \partial \bar{x}$. Here, $S_0$ is the stream function in the leading order. The solutions for the velocity easily follow from (4.6) and the boundary conditions, which read:

$$\hat{u}_0(\bar{y}) = -n\left(1 - e^{-\bar{y}/\delta}\right); \quad S_0 = \frac{\delta^2}{\delta_1^2}\left[\{\delta_1 + (\delta_1 - 1)\bar{y}\}e^{-\bar{y}} - \delta_1 e^{-\bar{y}/\delta_1}\right]\cos(\bar{x}) \tag{4.7}$$

### 4.2. O(Pe) solutions

The first order equations for charge distribution, potential and momentum can be deduced from (3.2 – 3.5), which read:

$$\bar{\nabla}^2 \bar{\rho}_e^{(1)} + 2\bar{\zeta}_0 \bar{\nabla}^2 \bar{\varphi}_1 = \bar{u}_0 \frac{\partial \bar{\rho}_e^{(0)}}{\partial \bar{x}} + \bar{v}_0 \frac{\partial \bar{\rho}_e^{(0)}}{\partial \bar{y}} \tag{4.8}$$

$$\bar{\nabla}^2 \bar{\varphi}_1 = -\frac{\bar{\rho}_e^{(1)}}{2\bar{\zeta}_0 \delta^2} \tag{4.9}$$

$$-\bar{\nabla}\bar{p}_1 + \nabla^2 \bar{\mathbf{v}}_1 + \beta\left(\bar{\nabla}^2 \bar{\varphi}_1 \bar{\nabla}\bar{\varphi}_0 + \bar{\nabla}^2 \bar{\varphi}_0 \bar{\nabla}\bar{\varphi}_1\right) - \bar{\nabla}^2 \bar{\varphi}_1 \mathbf{e_x} = 0; \quad \bar{\nabla} \cdot \bar{\mathbf{v}}_1 = 0 \tag{4.10}$$

These equations are subject to the following boundary conditions:

$$\frac{\partial \bar{\rho}_e^{(1)}}{\partial y} + 2\bar{\zeta}_0 \frac{\partial \bar{\varphi}_1}{\partial \bar{y}} = 0, \quad \bar{\varphi}_1 = 0 \text{ and } \mathbf{v}_1 = 0 \text{ at } \bar{y} = 0 \tag{4.11}$$

and at $\bar{y} \to \infty$, $\bar{\varphi}_1 \to 0, \bar{\rho}_e^{(1)} \to 0, \bar{v}_1 \to 0$ and $\frac{\partial \bar{u}_1}{\partial \bar{y}} \to 0$ \hfill (4.12)

We again note that the charge density and the potential at O(*Pe*) are independent of the velocities at O(*Pe*) and therefore, can be evaluated independently. Therefore, equations (4.8 – 4.9) can be solved by defining, $\bar{\rho}_e^{(1)} + 2\bar{\zeta}_0 \bar{\varphi}_1 = X_1(\bar{x}, \bar{y})$, where $X_1$ satisfies an equation of the form: $\bar{\nabla}^2 X_1 = \bar{u}_0 \frac{\partial \bar{\rho}_e^{(1)}}{\partial \bar{x}} + \bar{v}_0 \frac{\partial \bar{\rho}_e^{(1)}}{\partial \bar{y}} = Y_1^{(1)}(\bar{y})\sin(\bar{x}) + Y_2^{(1)}(\bar{y})\sin(2\bar{x})$. Expressions for $Y_1^{(1)}$ and $Y_2^{(1)}$ have been given in Appendix A. These are simply obtained by plugging in the leading order variables in the right hand side of equation (4.8). The boundary conditions for $X_1$ would simply read, $\partial X_1 / \partial Y(\bar{x}, \bar{y} = 0) = 0; \quad X_1(\bar{x}, \bar{y} \to \infty) = 0$. Therefore, we can easily infer that $X_1$ has a solution of the form:

$$X_1 = f_1(\bar{y})\sin(\bar{x}) + f_2(\bar{y})\sin(2x) \tag{4.13}$$

The equations for $f_k$'s can be easily deduced from the equation for $X_1$, as mentioned in the previous paragraph. These equations are of the following form: $(D^2 - k^2)f_k = Y_k^{(1)}$, $k = 1, 2$ and $D = d / d\bar{y}$. We can now find the solution for potential at O(*Pe*), by combining (4.9) and (4.13), which has the following form:

$$\bar{\varphi}_1 = g_1(\bar{y})\sin(\bar{x}) + g_2(\bar{y})\sin(2\bar{x}) \qquad (4.14)$$

We note that, in (4.15), $g_k$'s satisfy second order equations of the form, $(D^2 - \delta_k^{-2})g_k = -f_k/2\bar{\zeta}_0/\delta^2$, $k = 1, 2$, where $\delta_k = \left[\sqrt{k^2 + \delta^{-2}}\right]^{-1}$, while the functions $g_k$'s satisfy homogeneous boundary conditions. Equations of the aforementioned type can be solved analytically using a number of methods (for example variation of parameters). However, since the source terms in these equations are algebraically quite complicated, the solutions also turn out to be extremely difficult to handle. Therefore, we solve equations of this kind numerically, using a finite central difference scheme. The solutions from (4.8 – 4.9) can be enforced in (4.10) to evaluate the velocities at O($Pe$). These velocities have the following mathematical form: $\bar{u}_1 = \hat{u}_1(\bar{y}) + \partial S_1/\partial \bar{y}$; $\bar{v}_1 = -\partial S_1/\partial \bar{x}$, where $S_1$ is the stream function. The equation for $\hat{u}_1$ is given by:

$$D^2 \hat{u}_1 + \left(\beta/4\bar{\zeta}_0\delta^2\right)e^{-\bar{y}/\delta_1}f_1(\bar{y}) = 0 \qquad (4.15)$$

while the stream function ($S_1$) has the following form:

$$S_1 = \sum_{k=1}^{3} h_k^{(c)}(\bar{y})\cos(k\bar{x}) + \sum_{k=1}^{2} h_k^{(s)}(\bar{y})\sin(k\bar{x}) \qquad (4.16)$$

In (4.16) the functions $h_k$'s satisfy the fourth order ordinary equations of the form $(D^4 - 2k^2 D^2 + k^4)h_k^{(c)} = kV_k^{(s)} - DU_k^{(c)}$ and $(D^4 - 2k^2 D^2 + k^4)h_k^{(s)} = -DU_k^{(s)}$. The expression for the functions $U_k^{(c)}, V_k^{(c)}$... etc have been given in Appendix-A. We solve the foregoing fourth order equations numerically, using a simple central difference scheme, since the solutions are too complex to handle analytically. It is interesting to note that at O($Pe$), a net flow is generated because of the contribution from the term $f_1$ in (4.15). This term, as evident from (4.12), is an outcome of the interaction between the constant component of the surface potential (i.e., $n$) and the periodic component of the same. In a closed form confinement this would result in an alteration in the net throughput. However, since in the present study we are dealing with a semi-infinite region of liquid, the net velocity alters the free stream velocity at $\bar{y} \to \infty$ (discussed in more detail in section 4.4).

### 4.3. O($Pe^2$) solutions

The O($Pe^2$) equations can be deduced from (3.2 – 3.5), which read:

$$\bar{\nabla}^2 \bar{\rho}_e^{(2)} + 2\bar{\zeta}_0 \bar{\nabla}^2 \bar{\varphi}_2 = \bar{u}_0 \frac{\partial \bar{\rho}_e^{(1)}}{\partial \bar{x}} + \bar{v}_0 \frac{\partial \bar{\rho}_e^{(1)}}{\partial \bar{y}} + \bar{u}_1 \frac{\partial \bar{\rho}_e^{(0)}}{\partial \bar{x}} + \bar{v}_1 \frac{\partial \bar{\rho}_e^{(0)}}{\partial \bar{y}} \qquad (4.17)$$

$$\bar{\nabla}^2 \bar{\varphi}_2 = -\frac{\bar{\rho}_e^{(2)}}{2\bar{\zeta}_0 \delta^2} \qquad (4.18)$$

$$-\bar{\nabla}\bar{p}_2 + \bar{\nabla}^2 \bar{\mathbf{v}}_2 + \beta\left(\bar{\nabla}^2 \bar{\varphi}_2 \bar{\nabla}\bar{\varphi}_0 + \bar{\nabla}^2 \bar{\varphi}_0 \bar{\nabla}\bar{\varphi}_2 + \bar{\nabla}^2 \bar{\varphi}_1 \bar{\nabla}\bar{\varphi}_1\right) - \bar{\nabla}^2 \bar{\varphi}_2 \mathbf{e}_x = 0; \quad \bar{\nabla} \cdot \bar{\mathbf{v}}_2 = 0 \qquad (4.19)$$

Equations (4.17 – 4.19) are subjected to the following boundary conditions:

$$\frac{\partial \bar{\rho}_e^{(2)}}{\partial \bar{y}} + 2\bar{\zeta}_0 \frac{\partial \bar{\varphi}_2}{\partial \bar{y}} = 0, \; \bar{\varphi}_2 = 0 \text{ and } \mathbf{v}_2 = 0 \text{ at } \bar{y} = 0 \tag{4.20}$$

$$\text{at } \bar{y} \to \infty, \; \bar{\varphi}_2 \to 0, \bar{\rho}_e^{(2)} \to 0, \bar{v}_2 \to 0, \frac{\partial \bar{u}_2}{\partial y} \to 0 \tag{4.21}$$

The O($Pe^2$) solutions can be found in the exact same way as described for the O($Pe$) solutions in the previous subsection. The charge density is of the form: $\bar{\rho}_e^{(2)} = -2\bar{\zeta}_0 \bar{\varphi}_2 + X_2$, where $X_2$ has the following form, $X_2 = \sum_{k=1}^{3} f_k^{(2,c)}(\bar{y})\cos(k\bar{x}) + \sum_{k=1}^{4} f_k^{(2,s)}(\bar{y})\sin(k\bar{x}) + f_n^{(2)}(\bar{y})$. The functions $f_k^{(2,c)}$, $f_k^{(2,s)}$ etc satisfy analogous second order equations to the ones satisfied by $f_k$'s in (4.13), which are solved numerically using a central difference technique. Note that at O($Pe^2$), the charge density has a net or a non-periodic component, because of the contribution from the term $f_n^{(2)}(\bar{y})$. The potential can be easily deduced by combining (4.18) with the foregoing expression for charge distribution and reads:

$$\bar{\varphi}_2 = \sum_{k=1}^{3} g_k^{(2,c)}(\bar{y})\cos(k\bar{x}) + \sum_{k=1}^{4} g_k^{(2,s)}(\bar{y})\sin(k\bar{x}) + g_n^{(2)}(\bar{y}) \tag{4.22}$$

Again, in (4.22) the functions $g_k$'s satisfy analogous second order equations to the ones satisfied by the funcitons in (4.14). Finally, one can evaluate the velocity fields at O($Pe^2$), by combining (4.22) and (4.19). The velocity components can be expressed in the following form: $\bar{\mathbf{v}}_2 = \hat{u}_2(\bar{y})\mathbf{e_x} - \nabla \times S_2 \mathbf{e_z}$ ($S_2$ is the stream function). We again note that at O($Pe^2$), there exists a net non-periodic component of the $x$-velocity (here represented by $\hat{u}_2$), which alters the "far-field" or the "free-stream" velocity in the present flow geometry. We observe that $\hat{u}_2$ satisfies the following equation:

$$\frac{d^2 \hat{u}_2}{d\bar{y}^2} - \frac{2\beta\bar{\zeta}_0 g_1^{(2,s)}(\bar{y}) e^{-\frac{\bar{y}}{\delta_1}} \left(\delta_1^2 - \delta^2 + \delta^2 \delta_1^2\right) - 2\delta_1^2 f_n^{(2)} + 4\delta_1^2 \bar{\zeta}_0 g_n^{(2)} - \beta\delta_1^2 f_1^{(2,s)} e^{-\frac{\bar{y}}{\delta_1}}}{4\bar{\zeta}_0 \delta^2 \delta_1^2} = 0 \tag{4.23}$$

Equation (4.23) demonstrates that the component $\hat{u}_2$ is generated owing to the interactions between O(1), O($Pe$) and O($Pe^2$) potential as well as the non-periodic part of the charge density at O($Pe^2$). The stream function, on the other hand has the following form:

$$S_2 = \sum_{k=1}^{5} s_k^{(2,c)}(\bar{y})\cos(k\bar{x}) + \sum_{k=1}^{4} s_k^{(2,s)}(\bar{y})\sin(k\bar{x}) \tag{4.24}$$

In (4.24), the functions $s_k$'s satisfy analogous fourth order equations to the ones satisfied by the functions $h_k$'s in (4.16). Since these equations have algebraically complicated source terms, we do not explicitly mention the equations for the sake of brevity. The MATLAB and MAPLE files containing the expressions can be made available upon request to the authors.

*4.4 Free stream velocity*

Since in the present study we are only considering motion of a semi-infinite fluid body, one of the most important indicators of the electroosmotic mobility is the "far-field" or the free stream velocity. We define the "far-field" velocity in the following way:

$$\bar{u}_\infty = \lim_{\bar{y} \to \infty} \bar{u} \qquad (4.26)$$

It is very intuitive to infer from the solution (4.7) that the axially periodic components of the velocity vanish exponentially, as $\bar{y} \to \infty$ at any order of *Pe*. Therefore, the only contribution to the far field velocity comes from the non-periodic component of the solutions, i.e., from the "fully developed" $\bar{u}(\bar{y})$ type functions. Therefore, based on the solutions derived in sections 4.1 – 4.3, one can write the free stream velocity ($\bar{u}_{\infty,I}$) as:

$$\bar{u}_{\infty,I} = \lim_{\bar{y} \to \infty} \left( \hat{u}_0 + Pe\,\hat{u}_1 + Pe^2 \hat{u}_2 \right) + O(Pe^3) \qquad (4.27)$$

Therefore, we conclude that the inclusion of advection clearly alters the "free-stream" velocity, which stems from the net non-periodic components of flow at O(*Pe*) and O(*Pe*²). Thus, if a confinement in the form of a top plate is added to the system, the advection terms are expected to alter the net throughput in the channel. We later show that the O(*Pe*) and O(*Pe*²) terms always decrease the magnitude of the free stream velocity.

Finally, another important to note here is that, in executing the asymptotic analysis, we have considered terms uptoO (*Pe*²), while we have neglected terms of O($\bar{\zeta}_0^2$)/β (appearing in the last terms in (2.13) and (2.14)). It automatically implies that the condition $O(\bar{\zeta}_0^2) \ll O(Pe^2)$ has to be satisfied for the present analysis to remain valid. However, we later show that, even if this criterion is not met, asymptotic analysis still offers a fairly accurate approximation to the relevant variables, with errors in the tune of 1%.

## 5. Asymptotic Limit – II: Thin EDL limit (δ≪1)

As noted in section 3, the thin EDL limit requires application of matched asymptotic expansion, wherein, all the variables in the outer as well as inner layer are expanded using (3.8). We reiterate that in the present section, we assume the ionic Peclet number to have arbitrary values, without exceeding O(1) in magnitude (i.e., *Pe* ~ O(1) or less). We first present the scaled governing equations in the two separate layers and then proceed to solve them to obtain an asymptotic solution with corrections upto O($\delta^2$) terms.

*5.1. Equations in the Bulk (Outer layer)*

Starting from the non-dimensional equations in section 3, we first deduce the correct scales for the pertinent variables in the outer layer. To this end, we note that all the variables remain O(1) (non-dimensionally) in the outer layer. Furthermore, for ease of representation we define $\bar{\chi} = \bar{\rho}_e / 2\bar{\zeta}_0$. Here, we mention that, the variables in the outer layer will be represented with an "overbar". Enforcing the asymptotic expansion (3.8) in (3.2 – 3.5), one can write the governing equations in orders of $\delta$. These equations are as follows:

$$\overline{\chi}_i = 0 \tag{5.1}$$

$$\nabla^2 \overline{\varphi}_i = 0 \tag{5.2}$$

$$-\overline{\nabla}\overline{p}_i + \overline{\nabla}^2 \overline{\mathbf{v}}_i = 0; \quad \overline{\nabla} \cdot \overline{\mathbf{v}}_i = 0 \tag{5.3}$$

Equations (5.1 – 5.3) are valid for all values of $i$ ($i$ = 0, 1, 2, …). Therefore, the equations in the outer layer remain the same at every order of $\delta$. These equations are subject to the boundary conditions:

As $\overline{y} \to \infty$, $v_i$, $\chi_i$, $\varphi_i$ and $\partial \overline{u} / \partial \overline{y} \to 0$ \hfill (5.4)

We note that, we cannot expect these variables to satisfy the boundary conditions at the wall, since near the wall a separate layer exists. Therefore, we would expect the variables in the outer layer to match with those from the "*inner layer*" solutions, at the edge of the EDL. The matching conditions have been explicitly mentioned in the next subsection. We note from (5.1) that the bulk always remains electroneutral in the thin EDL limit and hence there are no net electrical body forces acting on the fluid in this layer.

### *5.2. Equations in the EDL (inner layer)*

In an effort to derive the governing equations inside the EDL, we first deduce the scales of the pertinent variables. To this end, we note that inside the EDL, $\tilde{y} \sim \delta, \tilde{x} \sim O(1)$, $\tilde{u} \sim O(1)$, $\tilde{\varphi} \sim O(1)$ and $\tilde{\chi} \sim O(1)$. Here we mention that inside the EDL, the non-dimensional variables will be noted by a "tilde" overhead. Following the aforementioned scales, the continuity equation suggests that we must have, $\tilde{v} \sim O(\delta)$. Furthermore, the *y*-momentum equation suggests that the pressure scales as, $\tilde{p} \sim O(\delta^{-2})$. Note that the same scale for pressure has been used in a handful of previous studies(Yariv et al. 2011; Schnitzer et al. 2012), in order to resolve the fluid dynamics within the thin EDL. We now introduce rescaled variables for the inner layer, in an effort to obtain the governing equations in this layer. To this end, we define, $\overline{u} \to U$, $\overline{\varphi} \to \tilde{\varphi}, \overline{\chi} \to \tilde{\chi}, \overline{x} \to X$, while the *y*-coordinate is rescaled as $y \to \delta Y$, the corresponding velocity rescaled as $\overline{v} \to \delta V$ and the pressure is rescaled as, $\overline{p} \to \delta^{-2} P$. Enforcing these new variables in (3.2 – 3.5) one obtains:

$$\frac{\partial^2 \tilde{\chi}}{\partial Y^2} + \frac{\partial^2 \tilde{\varphi}}{\partial Y^2} = \delta^2 \left[ Pe\left(U \frac{\partial \tilde{\chi}}{\partial X} + V \frac{\partial \tilde{\chi}}{\partial Y}\right) - \frac{\partial^2 \tilde{\chi}}{\partial X^2} - \frac{\partial^2 \tilde{\varphi}}{\partial X^2} \right] \tag{5.5}$$

$$\frac{\partial^2 \tilde{\varphi}}{\partial Y^2} + \tilde{\chi} = -\delta^2 \frac{\partial^2 \tilde{\varphi}}{\partial X^2} \tag{5.6}$$

$$\frac{\partial^2 U}{\partial Y^2} + \frac{\partial^2 \tilde{\varphi}}{\partial Y^2}\left(\beta \frac{\partial \tilde{\varphi}}{\partial X} - 1\right) - \frac{\partial P}{\partial X} = \delta^2 \left[ -\frac{\partial^2 U}{\partial X^2} - \frac{\partial^2 \tilde{\varphi}}{\partial X^2}\left(\beta \frac{\partial \tilde{\varphi}}{\partial X} - 1\right) \right] \tag{5.7}$$

$$\beta\frac{\partial^2\tilde{\varphi}}{\partial Y^2}\frac{\partial\tilde{\varphi}}{\partial Y}-\frac{\partial P}{\partial Y}=\delta^2\left[-\frac{\partial^2 V}{\partial Y^2}-\beta\frac{\partial\tilde{\varphi}}{\partial X}\frac{\partial^2\tilde{\varphi}}{\partial X^2}-\delta^2\frac{\partial^2 V}{\partial X^2}\right] \quad (5.8)$$

$$\frac{\partial U}{\partial X}+\frac{\partial V}{\partial Y}=0 \quad (5.9)$$

These equations are subject to the following boundary conditions at the wall:

At $Y = 0$, $U = V = 0$, $\tilde{\varphi} = n + \cos(X)$; and $\frac{\partial\tilde{\chi}}{\partial Y}+\frac{\partial\tilde{\varphi}}{\partial Y}=0$ \quad (5.10)

At the upper edge of the EDL, these variables have to be matched with their counterparts in the bulk, taking into account the scaling introduced earlier, for the inner layer. The matching conditions for different variables are as follows:

$$\lim_{Y\to\infty}\tilde{\chi}=\lim_{\bar{y}\to 0}\bar{\chi}=0; \quad \lim_{Y\to\infty}\tilde{\varphi}=\lim_{\bar{y}\to 0}\bar{\varphi}$$

$$\lim_{Y\to\infty}U=\lim_{\bar{y}\to 0}\bar{u}; \quad \lim_{Y\to\infty}\delta V=\lim_{\bar{y}\to 0}\bar{v}; \quad \lim_{Y\to\infty}\frac{1}{\delta^2}P=\lim_{\bar{y}\to 0}\bar{p}; \quad (5.11)$$

However, it is important to note here that there is one more constraint to take care of, namely, the overall charge balance in the EDL. It has been previously shown (Yariv et al. 2011; Schnitzer et al. 2012; Ajdari 2000) that advection inside the EDL drives the charges around and hence an additional constraint is required, which balances the overall charge and concentration within the EDL at steady state. Such conditions have already been derived by Schnitzer and coworkers (Schnitzer et al. 2012; Yariv et al. 2011) for flows related to streaming potential. Storey et al. (Storey et al. 2008) and Ramos et al. (Ramos et al. 2003) also derived similar conditions, when the surface potential is AC in nature. However, these conditions were either derived assuming uniform surface potential (Yariv et al. 2011), or, without the presence of charge advection. Following these studies, here we give a very similar overall charge balance condition, which is somewhat more general taking into account the non-uniform surface potential and ionic advection required in the present case. The details of the derivation have been given in Appendix-B. It then follows from Appendix-B that, for low surface potential, the non-dimensionalized and scaled charge balance condition, reads:

$$\lim_{\bar{y}\to 0}\frac{\partial\tilde{\varphi}}{\partial\bar{y}}=\delta\int_0^\infty\frac{\partial}{\partial X}\left\{PeU\tilde{\chi}-\frac{\partial\tilde{\chi}}{\partial X}-\frac{\partial\tilde{\varphi}}{\partial X}\right\}dY \quad (5.12).$$

This equation demonstrates that, although in the governing equation the advection terms are expected to have a first effect at $O(\delta^2)$, they explicitly appear in the boundary conditions at $O(\delta)$. We thus conclude that advection of ion in the leading order of $\delta$, leads to a non-zero outer potential, which is in sharp contrast to the classical Poisson-Boltzmann distribution. In fact, bulk potential drives a current into the EDL, in order to maintain overall charge balance at steady state. Non-zero outer potentials also appear, when the electrolyte solution is subjected to an AC surface potential. However, in cases of AC surface potential, the potential in the bulk turns out to be non-zero in the leading order of $\delta$, as opposed to the present case.

Our aim, in the present section is to solve (5.5 – 5.10), subject to the conditions (5.10 – 5.12), using an asymptotic expansion for the variables of the type (3.8).

### 5.3. Leading Order solutions

The leading order equations remain same as in the case of negligible advection. Nevertheless, for the sake of completeness, we briefly derive them here. In the leading order the inner layer equations read:

$$\frac{\partial^2 \tilde{\chi}_0}{\partial Y^2} + \frac{\partial^2 \tilde{\varphi}_0}{\partial Y^2} = \frac{\partial^2 \tilde{\varphi}_0}{\partial Y^2} + \tilde{\chi}_0 = 0 \quad (5.13)$$

$$\beta \frac{\partial^2 \tilde{\varphi}_0}{\partial Y^2} \frac{\partial \tilde{\varphi}_0}{\partial Y} - \frac{\partial P_0}{\partial Y} = \frac{\partial^2 U_0}{\partial Y^2} + \frac{\partial^2 \tilde{\varphi}_0}{\partial Y^2}\left(\beta \frac{\partial \tilde{\varphi}_0}{\partial X} - 1\right) - \frac{\partial P_0}{\partial X} = \frac{\partial U_0}{\partial X} + \frac{\partial V_0}{\partial Y} = 0 \quad (5.14)$$

These are subject to conditions (5.10 – 5.12) at the boundaries. The outer layer equations read:

$$-\overline{\nabla} \overline{p}_0 + \overline{\nabla}^2 \overline{\mathbf{v}}_0 = 0; \quad \overline{\nabla} \cdot \overline{\mathbf{v}}_0 = 0; \quad \nabla^2 \overline{\varphi}_0 = 0; \text{ and } \overline{\chi}_0 = 0 \quad (5.15).$$

These equations are subjected to the conditions (5.4) at large distances from the surface and the matching conditions (5.11) and (5.12). We now outline the general solution procedure for the foregoing equations at any order of $\delta$:

(*i*) First solve for the bulk potential $\tilde{\varphi}_i$ ($i$ = 0, 1, 2, …) subject (5.4) and (5.12). The condition (5.12) can be employed without the knowledge of charge and velocity distribution in the current order of $\delta$.

(*ii*) Next, solve for the potential and charge distribution in the inner layer, i.e., for $\tilde{\varphi}_i$ and $\tilde{\chi}_i$, from the inner layer equations, (5.5) and (5.6). Since, the outer layer potential as well as charge density at *i*-th order of $\delta$ is already known (note that $\overline{\chi}_i = 0$ for all *i*), one can successfully apply the matching conditions for the charge distribution and potential in the EDL. Therefore, in this step we employ the conditions (5.10) and (5.11).

(*iii*) Once the potential is known, plug it in the *y*-momentum equation (5.8) to determine the pressure. When we apply the matching condition for pressure, we note that both $P_0$ and $P_1$ should go to 0 in the limit $Y \gg 1$. This is simply because the matching condition for pressure suggests that there are no terms of order $O(\delta^{-2})$ and $O(\delta^{-1})$ in the outer solutions. Therefore, $P_0$ and $P_1$ must decay to zero at the edge of the EDL. We further note that, we must match $P_2$ with $\overline{p}_0$ and so on, as evident from the matching conditions (5.11). Hence, $\lim_{Y \to \infty} P_2$ will be known a-priori, from the leading order outer solutions. Similar trends also follow for pressures of higher orders in $\delta$.

(*iv*) Next, plug in the pressure in (5.7) to determine *U*. This *U* can be used in the continuity equation (5.9) to determine *V*. From the matching conditions (5.11), it is clear that the limiting values of *U* and *V* for $Y \gg 1$, would give us the boundary conditions for the outer layer, or, the bulk velocities.

(v) Use the aforementioned boundary conditions for the outer layer velocities to solve for the velocity field in the bulk and proceed to the next order of solutions.

We now apply these five steps to obtain the solutions in the leading order. To this end, we first note from (5.12) that, since, $\partial \tilde{\varphi}_0 / \partial \bar{y} = 0$ at $\bar{y} = 0$, we must have, $\bar{\varphi}_0 = 0$. Therefore, in the leading order, the bulk potential remains zero, as we have already discussed. The solutions to (5.13) for charge density and potential then read:

$$\tilde{\chi}_0 = -\tilde{\varphi}_0; \quad \tilde{\varphi}_0 = \left[n + \cos(X)\right] e^{-Y} \tag{5.16}$$

The solutions for the pressure reads, $P_0 = \beta \tilde{\varphi}_0^2 / 2$ and the velocities are given by:

$$U_0 = -\left[n + \cos(X)\right]\left(1 - e^{-Y}\right); \quad V_0 = \sin(X)\left(1 - Y - e^{-Y}\right) \tag{5.17}$$

While evaluating the X velocity, we have applied the criteria that $U_0$ must remain bounded as, $Y \gg 1$. We note that, the matching criteria (5.11) are automatically satisfied for potential, charge density and the pressure. On the other hand, the matching conditions for velocities in the outer layer give boundary conditions for the same:

$$\lim_{\bar{y} \to 0} \bar{u}_0 = \lim_{Y \to \infty} U_0 = -n - \cos(X); \quad \lim_{\bar{y} \to 0} \bar{v}_0 = \lim_{Y \to \infty} \delta V = 0 + O(\delta); \tag{5.18}$$

Equations (5.15) in the outer layer can now be solved subject to (5.4) and (5.18). The solutions can be obtained based on a stream function formulation: $\bar{\mathbf{v}}_0 = -n\mathbf{e}_x - \bar{\nabla} \times \bar{S}_0 \mathbf{e}_z$. The stream function has the expression: $\bar{S}_0 = -\bar{y} e^{-\bar{y}} \cos(\bar{x})$.

Therefore, in the leading order of $\delta$, the velocities satisfy the much celebrated Smoluchowski slip conditions, even in presence of ionic advection, provided the same is not too strong.

### 5.4. The O($\delta$) Solutions

The O($\delta$) equations in the outer layer are given by:

$$-\bar{\nabla}\bar{p}_1 + \bar{\nabla}^2 \bar{\mathbf{v}}_1 = 0; \quad \bar{\nabla} \cdot \bar{\mathbf{v}}_1 = 0; \quad \nabla^2 \bar{\varphi}_1 = 0; \quad \text{and} \quad \bar{\chi}_1 = 0 \tag{5.20}$$

The equations in the inner layer are given by:

$$\frac{\partial^2 \tilde{\chi}_1}{\partial Y^2} + \frac{\partial^2 \tilde{\varphi}_1}{\partial Y^2} = \frac{\partial^2 \tilde{\varphi}_1}{\partial Y^2} + \tilde{\chi}_1 = 0 \tag{5.21}$$

$$\beta \frac{\partial^2 \tilde{\varphi}_0}{\partial Y^2} \frac{\partial \tilde{\varphi}_1}{\partial Y} + \beta \frac{\partial^2 \tilde{\varphi}_1}{\partial Y^2} \frac{\partial \tilde{\varphi}_0}{\partial Y} - \frac{\partial P_1}{\partial Y} = 0 \tag{5.22}$$

$$\frac{\partial^2 U_1}{\partial Y^2} + \beta \left( \frac{\partial^2 \tilde{\varphi}_0}{\partial Y^2} \frac{\partial \tilde{\varphi}_1}{\partial X} + \frac{\partial^2 \tilde{\varphi}_1}{\partial Y^2} \frac{\partial \tilde{\varphi}_0}{\partial X} \right) - \frac{\partial^2 \tilde{\varphi}_1}{\partial Y^2} \frac{\partial P_1}{\partial X} = \frac{\partial U_1}{\partial X} + \frac{\partial V_1}{\partial Y} = 0 \tag{5.23}$$

These equations are subjected to homogeneous boundary conditions at the walls and far away from the surface along with the conditions (5.11 – 5.12). Here, we note that, at O($\delta$) the condition (5.12) reads:

$$\left.\frac{\partial \bar{\varphi}_1}{\partial \bar{y}}\right|_{\bar{y}=0} = \frac{d}{dX}\int_0^\infty Pe U_0 \tilde{\chi}_0 dY = -Pe\left[n\sin(\bar{x}) + \frac{1}{2}\sin(2\bar{x})\right] \tag{5.24}$$

Therefore, the solutions for $\bar{\varphi}_1$ reads: $\bar{\varphi}_1 = Pe\left[ne^{-\bar{y}}\sin(\bar{x}) + \frac{1}{4}e^{-2\bar{y}}\sin(2\bar{x})\right]$ (5.25).

Equation (5.25) shows that, if $Pe \ll 1$, or, in other words if, ionic advection is completely neglected, the bulk potential at O($\delta$) is identically zero. We further note that at this order, there is interaction between the axially varying and constant components of the velocity, giving rise to additional variation in the potential, as evident from the term proportional to $n$ in (5.25). The matching condition for charge density and potential in the EDL then read: for charge density, $\lim_{Y\to\infty}\tilde{\chi}_1 = 0$ and for potential, $\lim_{Y\to\infty}\tilde{\varphi}_1 = \lim_{\bar{y}\to 0}\bar{\varphi}_1 = Pe\left[n\sin(X) + \frac{1}{4}\sin(2X)\right]$.

Equations (5.21) can now be solved for $\tilde{\varphi}_1$ and $\tilde{\chi}_1$, subject to (5.10) and the foregoing matching criteria. The solution reads:

$$\tilde{\chi}_1 = -\tilde{\varphi}_1 + Pe\left[n\sin(X) + \frac{1}{4}\sin(2X)\right] \tag{5.26}$$

$$\tilde{\varphi}_1 = Pe\left[n\sin(X) + \frac{1}{4}\sin(2X)\right]\left(1 - e^{-Y}\right) \tag{5.27}.$$

The solutions (5.27) combined with (5.22) gives the pressure at O($\delta$) as: $P_1 = \beta\left(\frac{\partial \tilde{\varphi}_1}{\partial Y}\frac{\partial \tilde{\varphi}_0}{\partial Y}\right)$. We again note that the pressure $P_1$ does satisfy the required matching criteria, $\lim_{Y\to\infty}\delta^{-1}P_1 = 0$, on account of the fact that there are no O($\delta^{-1}$) terms for the pressure in the outer layer. The corresponding velocities can now be found from (5.23) in the following form:

$$U_1 = U_{1n}(Y) + \sum_{k=1}^3 U_{1c}^{(k)}(Y)\cos(kX) + \sum_{k=1}^2 U_{1s}^{(k)}(Y)\sin(kX) \tag{5.28}.$$

In (5.28), $U_{1n}(Y) = \frac{1}{2}n\beta Pe\left(1 - e^{-Y}\right)$; and $U_{1c}^{(k)} = c_{1,1}^{(k)}Y + c_{2,1}^{(k)}\left(1 - e^{-Y}\right)$; while, $U_{1s}^{(k)} = s_{1,1}^{(k)}Y + s_{2,1}^{(k)}\left(1 - e^{-Y}\right)$. The constants $c_{2,1}^{(k)}$ and $s_{2,1}^{(k)}$ are given by: $c_{2,1}^{(1)} = (1/4)Pe\beta(4n^2 + 1)$; $c_{2,1}^{(2)} = n\beta Pe$; $c_{2,1}^{(3)} = (1/4)Pe\beta$ and $s_{2,1}^{(1)} = nPe$; $s_{2,1}^{(2)} = (1/4)Pe$. The other constants have to be determined from the matching conditions. To this end, we note that, matching condition at O($\delta$), expressed in terms of inner variable (Nayfeh 2011; Bender & Orszag 1999), reads (since the leading order terms have already been matched, here we only write the O($\delta$) terms):

$$\lim_{Y\to\infty}(U_0 + \delta U_1) = \lim_{\bar{y}\to 0}(\bar{u}_0 + \delta \bar{u}_1) \Rightarrow \lim_{Y\to\infty}U_1 = \bar{u}_1|_{\bar{y}=0} + Y\left.\frac{\partial \bar{u}_0}{\partial \bar{y}}\right|_{\bar{y}=0} = \bar{u}_1|_{\bar{y}=0} + 2Y\cos(X) \tag{5.29}.$$

Equation (5.29) suggests that for matching we must have,

$$\bar{u}_1\big|_{\bar{y}=0} = \frac{1}{2}n\beta Pe + \sum_{k=1}^{3} c_{2,1}^{(k)} \cos(k\bar{x}) + \sum_{k=1}^{2} s_{2,1}^{(k)} \sin(k\bar{x}) \qquad (5.30)$$

And $c_{1,1}^{(1)} = 2;\ c_{1,1}^{(k)} = 0, k = 2,3;\ s_{1,1}^{(k)} = 0, k = 1,2$, (5.31).

Proceeding in the same way, one can determine the boundary condition for the *y*-velocity in the outer layer, which turns out to be:

$$\bar{v}_1\big|_{\bar{y}=0} = \sin(\bar{x}) \qquad (5.32).$$

Using the solutions for $U_1$, we can easily determine $V_1$. We have mentioned the expression for $V_1$ in Appendix-C. Note that for matching the *y*-velocity at order $O(\delta)$, we do not require to know $V_1$ explicitly. Finally, we solve for the velocities in the outer layer, wherein, $\bar{\mathbf{v}}_1 = \frac{1}{2}n\beta Pe\mathbf{e_x} - \bar{\nabla}\times\bar{S}_1\mathbf{e_z}$, $\bar{S}_1$ being the stream function at $O(\delta)$. The stream function has the form:

$$\bar{S}_1(\bar{x},\bar{y}) = \left[1 + \left(c_{2,1}^{(k)} + 1\right)\bar{y}\right]e^{-\bar{y}}\cos\bar{x} + \sum_{k=2}^{3} c_{2,1}^{(k)} \bar{y} e^{-k\bar{y}} \cos(k\bar{x}) + \sum_{k=1}^{2} s_{2,1}^{(k)} \bar{y} e^{-k\bar{y}} \sin(k\bar{x}) \qquad (5.33).$$

We note that, at $O(\delta)$ the *x*-velocity has a net non-periodic component, uniform in space. It is very clear that this component eventually leads to a drop in the free stream velocity and is proportional to all *n*, *β* and *Pe*. Therefore, we conclude that the dominant or the leading order effect of the EDL on free stream velocity in a semi-infinite domain in presence of ionic advection is to slow down the net fluid motion. However, we also note that the axial field strength, denoted by $\beta^{-1}$ has an exact opposite effect on the free stream velocity, since increase in the field strength (or, a decrease in *β*) leads to an increase in the "far-field" velocity. We further discuss about the free stream velocity in subsection 5.6.

### 5.5. The $O(\delta^2)$ Solutions

The $O(\delta^2)$ equations in the outer layer are given by:

$$-\bar{\nabla}\bar{p}_2 + \bar{\nabla}^2\bar{\mathbf{v}}_2 = 0;\ \bar{\nabla}\cdot\bar{\mathbf{v}}_2 = 0;\ \nabla^2\bar{\varphi}_2 = 0;\ \text{and}\ \bar{\chi}_2 = 0 \qquad (5.34).$$

The equations in the inner layer read:

$$\frac{\partial^2 \tilde{\chi}_2}{\partial Y^2} + \frac{\partial^2 \tilde{\varphi}_2}{\partial Y^2} = Pe\left[U_0 \frac{\partial \tilde{\chi}_0}{\partial X} + V_0 \frac{\partial \tilde{\chi}_0}{\partial Y}\right] \qquad (5.35)$$

$$\frac{\partial^2 \tilde{\varphi}_2}{\partial Y^2} + \tilde{\chi}_2 = -\frac{\partial^2 \tilde{\varphi}_0}{\partial X^2} \qquad (5.36)$$

$$\beta\left(\frac{\partial^2 \tilde{\varphi}_0}{\partial Y^2}\frac{\partial \tilde{\varphi}_2}{\partial Y} + \frac{\partial^2 \tilde{\varphi}_2}{\partial Y^2}\frac{\partial \tilde{\varphi}_0}{\partial Y} + \frac{\partial^2 \tilde{\varphi}_1}{\partial Y^2}\frac{\partial \tilde{\varphi}_1}{\partial Y} + \frac{\partial^2 \tilde{\varphi}_0}{\partial X^2}\frac{\partial \tilde{\varphi}_0}{\partial Y}\right) + \frac{\partial^2 V_0}{\partial Y^2} - \frac{\partial P_2}{\partial Y} = 0 \qquad (5.37)$$

$$\frac{\partial^2 U_2}{\partial Y^2} + \beta\left(\frac{\partial^2 \tilde{\varphi}_0}{\partial Y^2}\frac{\partial \tilde{\varphi}_2}{\partial X} + \frac{\partial^2 \tilde{\varphi}_2}{\partial Y^2}\frac{\partial \tilde{\varphi}_0}{\partial X} + \frac{\partial^2 \tilde{\varphi}_1}{\partial Y^2}\frac{\partial \tilde{\varphi}_1}{\partial X}\right)$$
$$-\frac{\partial^2 \tilde{\varphi}_2}{\partial Y^2}-\frac{\partial P_2}{\partial X} = -\frac{\partial^2 U_2}{\partial X^2} - \beta\frac{\partial^2 \tilde{\varphi}_0}{\partial X^2}\frac{\partial \tilde{\varphi}_0}{\partial X} + \frac{\partial^2 \tilde{\varphi}_0}{\partial X^2} \quad (5.38)$$

$$\frac{\partial U_2}{\partial X} + \frac{\partial V_2}{\partial Y} = 0 \quad (5.39).$$

These equations are subject to homogeneous conditions at the wall (no flux for ions, no slip for velocity and specified potential at the wall) and in the far field region. Additionally, they are also subjected to the matching conditions (5.11) and the overall charge balance condition (5.12). We further note that, the ionic Peclet number first explicitly appears in the governing equations at $O(\delta^2)$, which was our motivation for carrying out the asymptotic analysis till this order. The charge balance condition, at $O(\delta^2)$ becomes:

$$\left.\frac{\partial \bar{\varphi}_2}{\partial \bar{y}}\right|_{\bar{y}=0} + Y\left.\frac{\partial^2 \bar{\varphi}_1}{\partial \bar{y}^2}\right|_{\bar{y}=0} = \frac{d}{dX}\int_0^\infty Pe(U_0\tilde{\chi}_1 + U_1\tilde{\chi}_0)dY - \frac{d}{dX}\int_0^\infty\left(\frac{\partial \tilde{\chi}_1}{\partial X} + \frac{\partial \tilde{\varphi}_1}{\partial X}\right)dY \quad (5.40)$$

Plugging the expression for $\bar{\varphi}_1$ from (5.25) into (5.40), one deduces:

$$\left.\frac{\partial \bar{\varphi}_2}{\partial \bar{y}}\right|_{\bar{y}=0} = \sum_{k=1}^{3} e_c^{(k)}\cos(k\bar{x}) + \sum_{k=1}^{4} e_s^{(k)}\sin(k\bar{x}) \quad (5.41).$$

In (5.41), the constants are given by, $e_c^{(1)} = -Pe^2(n+1/8)$; $e_c^{(2)} = -(3/2)Pe^2 n$; $e_c^{(3)} = -(3/8)Pe^2$; $e_s^{(1)} = Pe^2\beta n(5/8+n^2/2) + 2nPe$; $e_s^{(2)} = Pe^2\beta(3n^2/2+1/4) + 2Pe$ and $e_s^{(3)} = (9/8)Pe^2\beta n$; $e_s^{(4)} = (1/4)Pe^2\beta$. The outer potential is then given by:

$$\bar{\varphi}_2 = \sum_{k=1}^{3} g_c^{(k)} e^{-k\bar{y}}\cos(k\bar{x}) + \sum_{k=1}^{4} g_s^{(k)} e^{-k\bar{y}}\sin(k\bar{x}); \quad g_c^{(k)} = -e_c^{(k)}/k; \quad g_s^{(k)} = -e_s^{(k)}/k \quad (5.42).$$

The matching condition for the EDL potential at this order reads (noting that $\bar{\varphi}_0 = 0$), $\lim_{Y\to\infty}\tilde{\varphi}_2 = \bar{\varphi}_2|_{\bar{y}=0} + Y(\partial\bar{\varphi}_1/\partial\bar{y})_{\bar{y}=0}$, which finally translates to:

$$\lim_{Y\to\infty}\tilde{\varphi}_2 = \sum_{k=1}^{3} g_c^{(k)}\cos(k\bar{x}) + \sum_{k=1}^{4} g_4^{(k)}\sin(k\bar{x}) - PeY\left[n\sin(X) + \frac{1}{2}\sin(2X)\right] \quad (5.43).$$

The most interesting thing to note from (5.43) is that, the EDL potential does not tend to a constant value as we approach the edge of the EDL. In sharp contrast to the potentials at the previous orders it shows a $a+bY$ type variation for $Y \gg 1$. Therefore, solutions for EDL potential from (5.35 – 5.36) must demonstrate the foregoing asymptotic behavior for large $Y$ values. We now solve (5.35) and (5.36) to obtain the solutions for $\tilde{\chi}_2$ and $\tilde{\varphi}_2$. These solutions read (note that, we must have $\lim_{Y \to \infty} \tilde{\chi}_2 = 0$):

$$\tilde{\chi}_2 = -\tilde{\varphi}_2 + \sum_{k=1}^{3} g_c^{(k)} \cos(k\bar{x}) + \sum_{k=1}^{4} g_4^{(k)} \sin(k\bar{x}) - Pe\left[n\sin(X) + \frac{1}{2}\sin(2X)\right]\left[Y + (2+Y)e^{-Y}\right]$$

(5.44)

$$\tilde{\varphi}_2 = \sum_{k=1}^{4} \hat{\varphi}_{s,2}^{(k)}(Y)\sin(kX) + \sum_{k=1}^{3} \hat{\varphi}_{c,2}^{(k)}(Y)\cos(kX)$$

(5.45).

In (5.45), the various functions are: $\hat{\varphi}_{s,2}^{(1)} = g_s^{(1)}(1-e^{-Y}) - nPe\left[Y + (5/4)Ye^{-Y} + (1/4)Y^2 e^{-Y}\right]$, $\hat{\varphi}_{s,2}^{(2)} = g_s^{(2)}(1-e^{-Y}) - Pe\left[(1/2)Y + (1/8)e^{-Y}(5Y + Y^2)\right]$, $\hat{\varphi}_{s,2}^{(k)} = g_s^{(k)}(1-e^{-Y})$, $k = 3,4$; while, $\hat{\varphi}_{c,2}^{(1)} = g_c^{(1)}(1-e^{-Y}) - (1/2)Ye^{-Y}$, $\hat{\varphi}_{c,2}^{(k)} = g_c^{(k)}(1-e^{-Y})$, $k = 2,3$. We note that the EDL potential automatically satisfies the $Y$ variation in the asymptotic limit (5.43), for large $Y$. Proceeding as outlined in section 5.3, we can now determine the $O(\delta^2)$ corrections to the inner layer pressure and velocities from (5.37 – 5.39). To this end, we note that, at the current level of approximation, the pressure must satisfy the matching criteria: $\lim_{Y \to \infty}\left(\frac{1}{\delta^2}P_0 + \frac{1}{\delta}P_1 + P_2\right) = \lim_{y \to 0} \bar{p}_0$, which finally becomes, $\lim_{Y \to \infty} P_2 = -2\sin(X)$. Note that, the leading order pressure in the outer layer can be easily determined from the leading order solutions for the velocities. The $O(\delta^2)$ correction to the pressure thus reads:

$$P_2 = \beta\left(\frac{\partial \tilde{\varphi}_2}{\partial Y}\frac{\partial \tilde{\varphi}_0}{\partial Y} + \frac{1}{2}\left(\frac{\partial \tilde{\varphi}_1}{\partial Y}\right)^2 + \int^Y \frac{\partial^2 \tilde{\varphi}_0}{\partial X^2}\frac{\partial \tilde{\varphi}_0}{\partial Y}dY\right) + \frac{\partial V_0}{\partial Y} - \sin(X)$$

(5.46).

The $O(\delta^2)$ corrections to the velocity is given by:

$$U_2 = \sum_{k=1}^{4} U_{2,s}^{(k)}(Y)\sin(kX) + \sum_{k=1}^{5} U_{2,c}^{(k)}(Y)\cos(kX) + U_{2n}(Y)$$

(5.47).

The various functions in (5.47) are of the form, $U_{2,c}^{(1)} = W_c^{(1)}(Y) + c_{1,2}^{(1)}Y + c_{2,2}^{(1)} - (3/2)Y^2$, $U_{2,c}^{(k)} = W_c^{(k)}(Y) + c_{1,2}^{(k)}Y + c_{2,2}^{(k)}$, $(k = 2-5)$; $U_{2,s}^{(k)} = W_s^{(k)}(Y) + s_{1,2}^{(k)}Y + s_{2,2}^{(k)}$, $(k = 1-4)$ and $U_{2,n}^{(k)} = W_n^{(2)}(Y) + c_{n,1}^{(2)}Y + c_{n,2}^{(2)}$. The constants $c_{2,2}^{(k)}, s_{2,2}^{(k)}$ and $c_{n,2}^{(2)}$ can be determined from the no-slip boundary conditions at the wall. The other constants, i.e., $c_{1,2}^{(k)}, s_{1,2}^{(k)}, c_{n,1}^{(1)}$ etc., have to be determined from the matching conditions with the bulk velocity at the edge of the EDL. From the first set of constants, we only mention the expression for $c_{n,2}^{(2)}$, which is given by:

$$c_{n,2}^{(2)} = -\beta n Pe\left(\frac{19}{8} + \frac{1}{4}n^2\beta Pe + \frac{5}{16}Pe\beta\right).$$ The expressions for the others have been mentioned in Appendix-C. Additionally, expressions for the functions $W_c^{(k)}, W_s^{(k)}$... etc., have also been

mentioned in the Appendix-C. Using the X-velocity, we can easily evaluate the Y-velocity, from the continuity equation. However, $V_2$ is not required for calculating the corrections to the bulk velocities in $O(\delta^2)$ and hence we do not explicitly mention the expression for $V_2$. Finally, to solve the outer layer velocities, from (5.34), we first employ the matching condition at $O(\delta^2)$. Following the norms of matching as outlined in section 5.4, we deduce:

$$\lim_{Y \to \infty} \left( U_0 + \delta U_1 + \delta^2 U_2 \right) = \lim_{\bar{y} \to 0} \left( \bar{u}_0 + \delta \bar{u}_1 + \delta^2 \bar{u}_2 \right) \tag{5.48}$$

$$\Rightarrow \lim_{Y \to \infty} U_2 = \bar{u}_2 \big|_{\bar{y}=0} + Y \frac{\partial \bar{u}_1}{\partial \bar{y}} \bigg|_{\bar{y}=0} + \frac{1}{2} Y^2 \frac{\partial^2 \bar{u}_0}{\partial \bar{y}^2} \bigg|_{\bar{y}=0} \tag{5.49}$$

Plugging in the expressions for the relevant variables in (5.49), we finally obtain:

$$c_{2,n}^{(2)} + \sum_{k=1}^{5} \left( c_{1,2}^{(k)} Y + c_{2,2}^{(k)} \right) \cos(kX) + \sum_{k=1}^{4} \left( s_{1,2}^{(k)} Y + s_{2,2}^{(k)} \right) \sin(kX) - \frac{3}{2} Y^2 \cos(X) =$$

$$\bar{u}_2 \big|_{\bar{y}=0} + Y \left[ -\left(2c_{1,2}^{(1)} + 1\right) \cos(\bar{x}) - \sum_{k=2}^{3} 2c_{2,1}^{(k)} k \cos(k\bar{x}) - \sum_{k=1}^{2} 2s_{2,1}^{(k)} k \sin(k\bar{x}) \right] - \frac{3}{2} Y^2 \cos(\bar{x})$$

(5.50)

Equation (5.50) suggests that, for matching we must require,

$$\bar{u}_2 \big|_{\bar{y}=0} = c_{n,2}^{(2)} + \sum_{k=1}^{5} c_{2,2}^{(2)} \cos(k\bar{x}) + \sum_{k=1}^{4} s_{2,2}^{(2)} \sin(k\bar{x}) \tag{5.51}$$

The other constants are given by: $c_{1,2}^{(1)} = -\left(2c_{2,1}^{(1)} + 1\right)$; $c_{1,2}^{(k)} = -2kc_{2,1}^{(k)}$ $(k=2,3)$; $c_{1,2}^{(k)} = 0$ $(k=4,5)$, $s_{1,2}^{(k)} = -2ks_{2,1}^{(k)}$ $(k=1,2)$; $s_{1,2}^{(k)} = 0$ $(k=3,4)$. Following a very similar approach for matching the y-velocity, one can write,

$$\bar{v}_2 \big|_{\bar{y}=0} = -\sum_{k=1}^{3} kc_{2,1}^{(k)} \sin(k\bar{x}) + \sum_{k=1}^{2} ks_{2,1}^{(k)} \cos(k\bar{x}) \tag{5.52}$$

The outer layer or, the bulk velocity can now be solved for, based on the conditions (5.50) and (5.52). As usual, to this end, we define a stream function such that the velocity is given by: $\bar{\mathbf{v}}_2 = -\beta n Pe \left( \frac{19}{8} + \frac{1}{4} n^2 \beta Pe + \frac{5}{16} Pe\beta \right) \mathbf{e_x} - \bar{\nabla} \times \bar{S}_2 \mathbf{e_z}$, where $\bar{S}_2$ is the correction to the stream function at $O(\delta^2)$. The stream function has the following expression:

$$\bar{S}_2 = \sum_{k=1}^{5} S_{2,c}^{(k)} \cos(k\bar{x}) + \sum_{k=1}^{4} S_{2,s}^{(k)} \sin(k\bar{x}) \tag{5.53}$$

In (5.53), the various functions are given by, $S_{2,c}^{(k)} = \left[ \left( c_{2,2}^{(k)} - kc_{2,1}^{(k)} \right) \bar{y} - c_{2,1}^{(k)} \right] e^{-k\bar{y}}$ $(k=1-3)$, $S_{2,c}^{(k)} = c_{2,2}^{(k)} \bar{y} e^{-k\bar{y}}$ with $(k=4,5)$, $S_{2,s}^{(k)} = \left[ \left( s_{2,2}^{(k)} - ks_{2,1}^{(k)} \right) \bar{y} - s_{2,1}^{(k)} \right] e^{-k\bar{y}}$ with $(k=1-2)$ and

$S_{2,s}^{(k)} = s_{2,2}^{(k)} \bar{y} e^{-k\bar{y}}$ $(k = 3, 4)$. We note that at O($\delta^2$) there is a again a correction to the "free-stream" velocity, as given by the net non-periodic component of the *x*-velocity. We discuss more on this, in the next subsection. We note that, at this level of approximation, the corrections to the non-periodic component of velocity comes from *Pe* explicitly appearing in the governing equations, as well as its influence through the overall charge balance condition. One can of course, carry on similar analysis for higher orders of *δ*, following the exact same procedure outlined in section (5.3). However, at these levels of corrections, the algebra quickly becomes very complex and intractable. We finally note that, both the asymptotic limits, which have been discussed in the present study, i.e., the one with *Pe*≪1 (weak advection limit) and the thin EDL limit, show that presence of ionic advection eventually alters the net fluidic transport, adding to the periodicity of the flow.

### 5.6. The free Stream Velocity

Based on the solutions for the velocities till O($\delta^2$), we can very easily calculate the "free stream" velocity $\left(\bar{u}_{\infty,II}\right)$, in presence of ionic advection. The expression goes as follows:

$$\bar{u}_{\infty,II} \sim -n + \delta\left(\frac{1}{2}n\beta Pe\right) - \delta^2\left[\beta n Pe\left(\frac{19}{8} + \frac{1}{4}n^2\beta Pe + \frac{5}{16}Pe\beta\right)\right] + \ldots \text{ as } \delta \to 0 \qquad (5.54)$$

We specifically use the asymptotic sign here, since the series mentioned above, in the powers of *δ* is likely to diverge, as is expected from singular perturbation (Bender & Orszag 1999). Therefore, the above expression for the free stream velocity is likely to hold only for vanishingly small *δ*. This expression shows that, the leading or, the most dominant effect of EDL thickness, in presence of strong applied axial electric field and ionic advection, is to slow down the net fluid movement. However, the O($\delta^2$) correction to $\bar{u}_{\infty,II}$ is seen to increase (in magnitude) the same. Note that, the O($\delta$) correction always have the opposite sign and the O($\delta^2$) corrections always have the same sign as that of the leading order velocity, even if the sign of *n* is altered. This is expected, since, altering the sign of *n* should not change the magnitude of the free stream velocity. We finally reiterate that, the axial field strength (here denoted by $\beta^{-1}$) can be noted to influence the free stream velocity in the opposite manner to that of advection strength, denoted by *Pe*.

As a final note, we again mention that we have neglected the O($\bar{\zeta}_0^2$) contributions to the variables in the present analysis, although we have included terms, upto O($\delta^2$). Therefore, we would expect the present analysis to offer accurate results, if $\delta^2 \gg \bar{\zeta}_0^2$. However, in the forthcoming sections, we show that, even if the foregoing condition is not satisfied, the asymptotic analysis still offers a fairly accurate solution, within its range of validity, with errors in the tune of 1%, or even less.

### 6. Numerical Solutions

Asymptotic analysis, executed in the previous sections, only offer approximate solutions with a limited range of validity. Therefore, in an effort to validate our analytical predictions and to explore the effects of advection for higher values of *Pe*, beyond the linearized limit, we attempt to solve the full set of governing equations, i.e., equations (2.13 – 2.17), subject to the conditions, (2.18 – 2.22) numerically. To this end, we have used a finite element (Hughes 2012) based framework to discretize the related governing equations. The system of discretized algebraic equations were subsequently solved using a Damped Newton method (Deuflhard 1974). In this method, the system of non-linear equations are linearized around a proper initial guess value, which are then discretized and solved to find the Newton step, using a linear solver. Here, we have employed the MUMPS (Multifrontal Massively Parallel Sparse) direct linear solver (Paszyński et al. 2010; Operto et al. 2007). The corrected value of the initial guess is computed in the following way: $U_1 = U_0 + \lambda \Delta U$, where $U_0$ is the initial guess, $\lambda$ is the damping factor and $\Delta U$ is the Newton step. Next, the error $E_r$ is computed from $U_1$. If the value of $E_r$ is smaller than its corresponding value of the previous step, the same steps are repeated to find the next correction. Otherwise the value of the damping factor is decreased to reduce the error. The whole computation stops, when the estimated error goes below the pre-specified relative tolerance. In the present study, we have set the relative tolerance at 0.001.

For our numerical computation, we chose a domain of dimension $2\pi \times H$ where the value of $H$ varies based on the EDL thickness. Since we are considering a semi-infinite region in the present study, the height of the domain should be large enough so that all the velocity gradients vanish as well as the potential reaches zero, when $y = H$. This is accomplished by taking $H$ at least 10 times larger than the EDL thickness. For $\delta < 1$, we have taken $H = 10$, while for $\delta > 1$, a value of $H = 20$ was chosen to meet the boundary conditions (2.22), i.e., the "far field" conditions. The whole domain was divided into rectangular grids of non-uniform density. The length along the *x* axis was divided into 30 uniformly spaced grid points, while along the *y* axis the grid points were non-uniformly placed. More specifically, a larger number of grid points were taken near the wall, to ensure that the sharp variations in the relevant variables inside the EDL were properly captured. To this end, the grid sizes were increased continuously from the wall ($\bar{y} = 0$), governed by a pre-defined ratio between the first and the last grid. This ratio was taken in two different ways, depending on the size of the EDL as compared to the modulation wavelength, namely, *Arithmetic Sequence*, where grid size was increased linearly and *Geometric Sequence*, where the grid sizes were increased exponentially. The latter was chosen for thinner EDLs. For example, for $\delta = 2$ ($H = 20$), 100 grid points were taken along the *y* axis, with the aforementioned ratio being 0.2, while the grid size was increased arithmetically. On the other hand, for $\delta = 0.1$ ($H = 10$), 80 grid points were taken along the *y* axis, the ratio being 0.02, while the grid size was increased geometrically. For thinner EDL thicknesses, especially for, $\delta < 0.05$, we have taken 200 or more grid points along the *y*-axis. As an example we mention that, for $\delta = 0.03$, we have taken 200 grid points along *y*-axis, while the grid sizes were increased geometrically, with the element ratio being 0.005. Figure 2 shows a sample grid distribution for $\delta = 0.3$ and $H = 10$. Here, we have taken 80 grid points along the *y* axis, while the grid sizes were increased geometrically, with an element ratio 0.08. As mentioned earlier, the number of grids along the *x* axis was taken as 30, for this case.

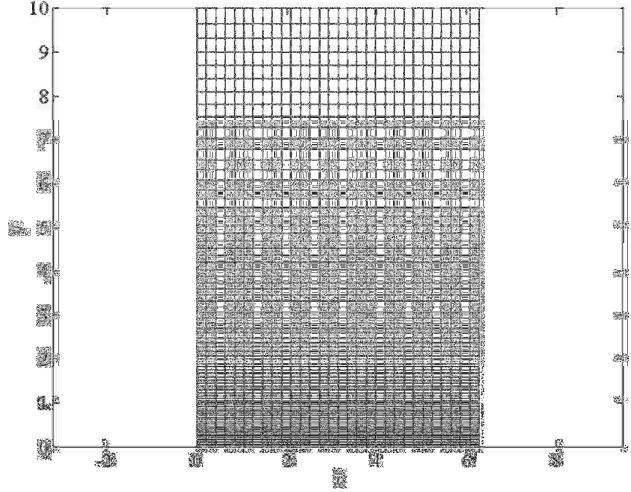

Figure 2: A sample mesh distribution used for simulation. This mesh has been used for $\delta = 0.3$, while the grid size in the $y$ direction was increased through a Geometric ratio of 0.08. The domain height was taken as $H = 10$ and width $W = 2\pi$. A total of 80 elements were taken along the $y$ axis, while 30 equal sized elements were taken along the $x$ axis.

We have defined the "far field" velocity based on the numerical solutions in the following way:

$$\bar{u}_\infty = \frac{1}{2\pi} \int_0^{2\pi} \bar{u}\left(\bar{y} = H, \bar{x}\right) d\bar{x} \tag{6.1}$$

## 7. Results and Discussions

In this section, we would focus on the variations in three separate aspects of the flow, which are, (a) the flow pattern or the flow dynamics (through streamline patterns), (b) the potential and charge distribution, which are coupled with the flow through the PNP equations and (c) the "free stream" velocity, $\bar{u}_\infty$. Before, we move towards the detailed results, it is important to point out the parameters, which are likely to influence the flow dynamics as well as the potential and charge distribution. Following the asymptotic analysis, we note these parameters to be: (i) The Péclet number ($Pe$), (ii) The axial field strength, represented by $\beta$, (iii) The non-dimensional EDL thickness ($\delta$) and, (iv) The parameter $n$. Since, in the present study our aim is to investigate the effect of ionic advection on the flow beyond the weak field limit, we would focus mostly on the first three of the four parameters mentioned above.

### 7.1. Variation in Potential

We begin our discussion with figure 3.(a) and (b), where the variations in the electrostatic potential $\varphi$, as a function of $y$ have been depicted at a particular $x$-location ($x = 2.138$ for (a) and $x = 2.404$ for (b)), for different values of Pèclet numbers, while the other parameters have been mentioned in the caption. Figure 3.(a) depicts the variation in potential, in the limit of thin EDL ($\delta = 0.05$), where numerical solutions along with the analytical solutions based on thin EDL limit from (section 5) have been plotted for $Pe = 0, 1, 2$ and 5. We first note that, reasonably good agreement is observed between the two, till $Pe \sim 2$.

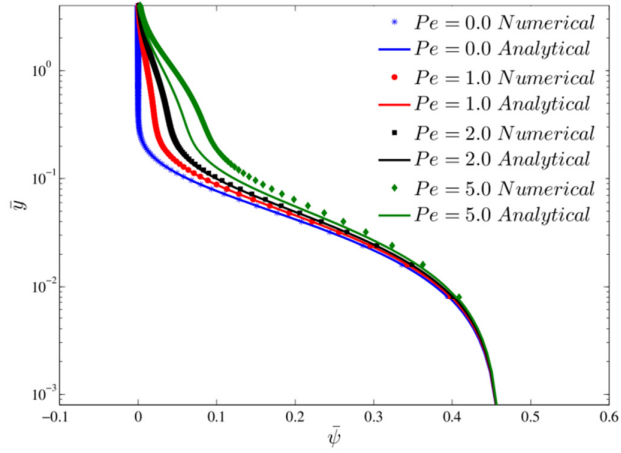

(a)

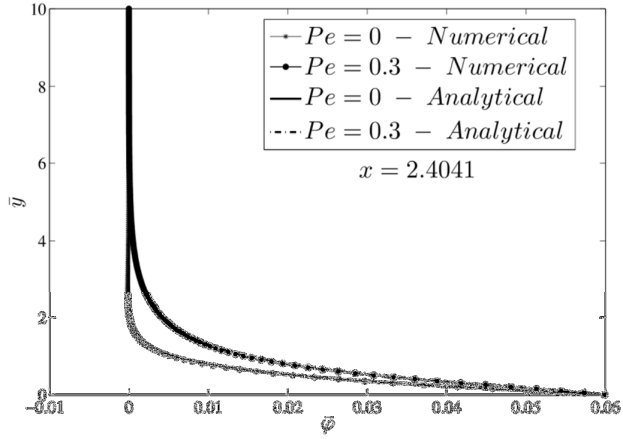

(b)

Figure 3:(a) Plot of $\bar{\varphi}_{uniform}$ vs $y$, for four different values of $Pe = 0, 1, 2$ and $5$, while the values of the other parameters are given by: $\zeta = 0.1$, $n = 1$, $\delta = 0.05$, $\beta = 1$, $x = 2.138$. We have plotted the numerical solutions along with the analytical solutions from the matched asymptotics ($Pe \sim O(1)$ and small $\delta$, section 5) and they show very good agreement for $Pe = 0$ (trivial case), 1 and 2. The agreement is relatively poor for $Pe = 5$, although the overall qualitative trend is rightly captured. (b) Plot of $\bar{\varphi}$ vs $\bar{y}$ for two different values of $Pe = 0$ (without any advection) and 0.3, depicted at $x$ location $x = 2.404$. Here, numerical solutions as well as the analytical solutions based on regular perturbation (small $Pe$, section 4) have been plotted. The other relevant parameters have the following values: $\bar{\zeta}_0 = 0.1$, $n = 0.8$, $\beta = 1$ and $\delta = 0.3$. Reasonably good agreement is observed between the numerical and analytical solutions.

The agreement is relatively poor for $Pe = 5$, although the qualitative trend is accurately represented by the analytical solution. Here is it important to mention that, we have plotted the uniform, or the combined expansion of $\bar{\varphi}$, constructed from inner and outer layer solutions, which is valid throughout the domain of interest. A second point to note from the present figure is that, we indeed see non-zero potential in the outer layer, when advection is taken into account. This is in sharp contrast to the case of $Pe = 0$, as clearly seen the figure. As expected, the outer potential is larger for larger advection strength. We further observe that the solution for potential in $O(\delta^2)$ varies as $\sim Pe^2$. Since for $Pe = 5$, $Pe^2 = 25 > O(1)$, we naturally do not expect the solution for $Pe = 5$ to match closely with the numerical solutions.

However, the qualitative trend in the variation of potential is in fact rightly represented by the analytical solution, with a tolerable error percentage. The reason for a non-zero outer potential can be very easily explained based on the overall charge balance condition (5.12). Since the advection of ions give rise to additional ionic fluxes, an outer potential has to be induced, which drives a net current into the EDL, in order to conserve the total amount of the charges inside it.

Figure 3.(b) shows a similar variation in the potential as a function of $y$ coordinate for thicker EDL ($\delta = 0.3$) and smaller Peclet numbers ($Pe = 0$ and 0.3). Along with the numerical solutions, we have also plotted the analytical solutions as deduced in the low $Pe$ regime (section 4, $Pe \ll 1$; from here onwards, we would refer to this solution as the weak advection regime), or, the weak advection regime. We again note that reasonably good agreement is observed between the two solutions. The exact same trend, as noted in the previous figure is followed here, i.e., potential shows a non-zero value, well beyond the EDL. The relative difference in the potentials between the two cases (i.e., $Pe = 0$ and 0.3) is larger here, as compared to the thin EDL case. This is simply because a thicker EDL would naturally induce higher bulk potentials, as indicated by the analytical solutions in the thin EDL limit.

### 7.2. Free Stream velocity

### 7.2.1 Effects of Ionic advection

Figure 5 depicts the variation in $\bar{u}_\infty$ as a function of $\delta$ (non-dimensional EDL thickness), for four different values of $Pe = 0.2, 0.3, 1$ and 2, therefore, revealing the effects of ionic advection at a given applied field strength (constant $\beta$). The values of the other relevant parameters have been mentioned in the caption. We have plotted the singular perturbation solutions from the thin EDL limit (section 5, eq. 5.54), for low values of $\delta$, for all four choices of $Pe$ values. For $Pe = 0.2$ and 0.3, asymptotic solutions based on weak advection regime have also been plotted for all values of EDL thickness. We first observe that the analytical solutions from the thin EDL limit resemble the numerical solutions with reasonably good accuracy for all values of $Pe$, in the regime of low EDL thicknesses. On the other hand, the analytical solution from the weak advection limit (section 4, eq. 4.27) closely follows the numerical results for low values of $Pe$, albeit for all EDL thicknesses. This, in essence demonstrates the effectiveness of the two separate asymptotic limits in determining the approximate solution to the problem in hand. It is important to note here that, in the present figure, the condition on the magnitude of zeta potential is not strictly satisfied in all the cases (for example, for $Pe = 0.2$, the condition $O(\bar{\zeta}_0^2) \ll O(Pe^2)$ condition is not satisfied). However, still reasonably good quantitative agreement is observed between the analytical and numerical solutions. There are several interesting points to note from the present figure. First, the free stream velocity decreases as $Pe$ is increased. We reiterate that, without ionic advection and considering weak field limit, the free stream velocity in this case is -2. The decrement in the free stream velocity with increasing $Pe$ can be very easily explained based on the analytical solutions in the thin EDL limit (section 5). From the expression (5.54) we can conclude that the effect of Peclet number first enters (for thin EDLs) in the $O(\delta)$ solutions. This is thus the most dominant effect of $Pe$, provided it does not exceed O(1) magnitude. Since this dominant effect leads to a decrement in the free stream velocity, as evident from (5.54), we naturally expect the free stream velocity to go down, as $Pe$ is increased. From a physical point of view, we note that the recirculation rolls induced in the flow, tend to drive the charges along with them. This leads to a change in the charge distribution in the fluid, which finally results in a decrease in the free stream velocity. A larger $Pe$ signifies larger velocity, which naturally causes a larger change in $\bar{u}_\infty$, as noted from the present figure. Second, we note that, for all $Pe$ values, there occurs a $\delta$, for which this decrease in $\bar{u}_\infty$ is maximum. Interestingly, the same trend is exactly reflected in the asymptotic solutions for weak advection limit as well, with the value of $\delta$ corresponding to the maximum decrement coinciding for analytical as well as numerical solutions. We further note that this value of $\delta$ is almost same for all values of $Pe$, i.e., close to 0.3. As the EDL thickness increases, the decrement in $\bar{u}_\infty$ almost vanishes. This unusual variation can be explained from the fact that, for low EDL thickness (smaller $\delta$) most of the charges reside very close to the wall, where the velocities are low, because of no-slip and no-penetration boundary conditions. Therefore, the fluid flow cannot recirculate the charges very effectively for thin EDLs. For higher EDL thicknesses, the charges are diffused over a larger length, away from the wall. However, in such cases, the velocities and the charge density inside the EDL are typically small, which again diminishes the recirculation of charges. Therefore, one can infer that, there exists a certain EDL thickness, for which the charge distribution is such that, maximum charge recirculation is brought about by the corresponding fluid velocities, thereby causing the maximum decrement in the free stream velocity.

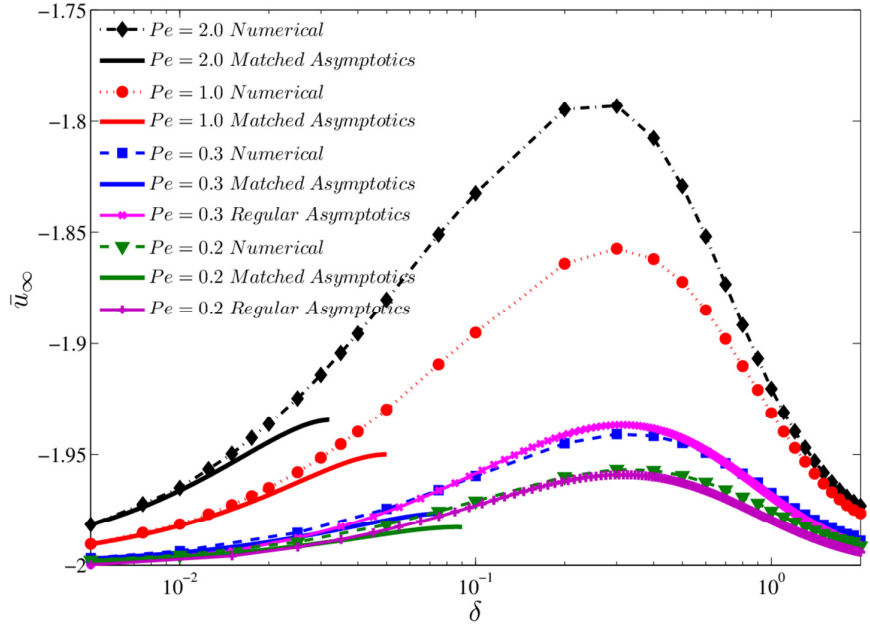

Figure 4: Plot of $\bar{u}_\infty$ vs $\delta$, for different values of $Pe$ = 0.2, 0.3, 1 and 2. The values of the other parameters are given by: $\zeta_0$ = 0.3, $\beta$ = 2.0 and $n$ = 2. We have plotted the numerical solutions along with the analytical solutions, for comparison. For $Pe$ = 0.2 and 0.3, numerical, analytical solutions from both regular asymptotics (for small $Pe$) and matched asymptotics (for small $\delta$) have been plotted and they show reasonably good agreement. For $Pe$ = 1.0 and 2.0, only the solutions from the matched asymptotics were plotted. These show good agreement with numerical solutions in the limit of thin EDL.

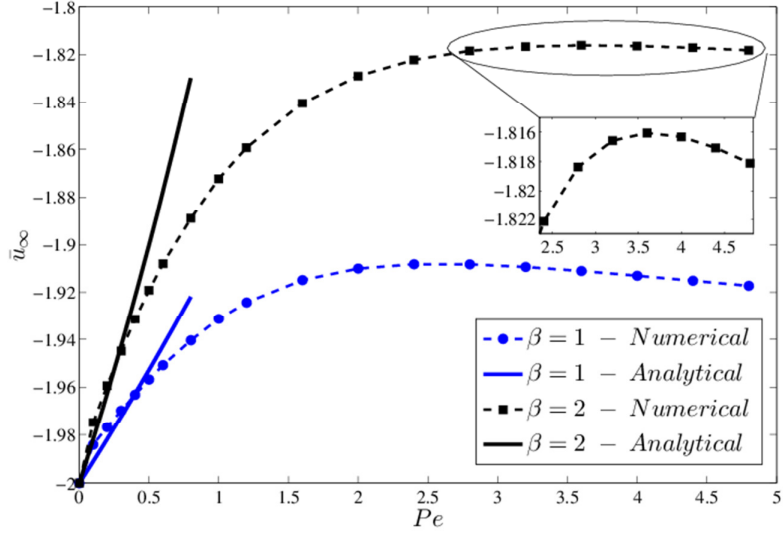

(a)

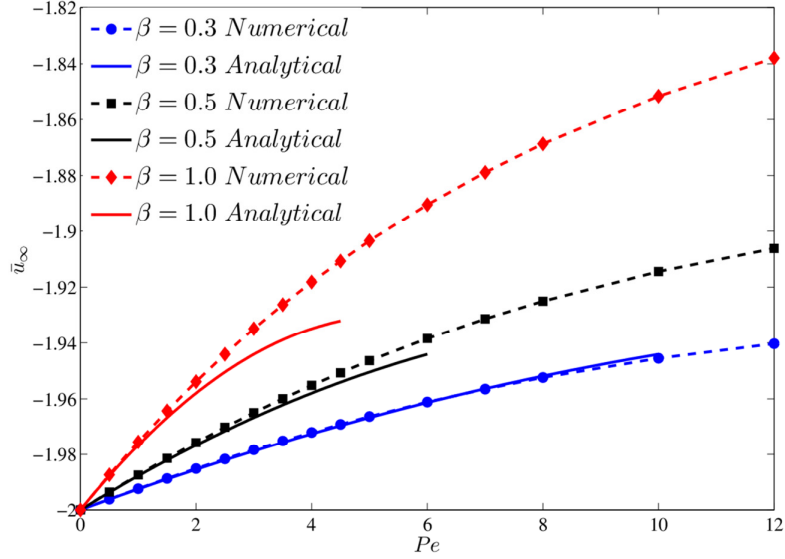

(b)

Figure 5: (a) $\bar{u}_\infty$ vs $Pe$, for two different values of $\beta = 1$ and 2. Numerical solutions have been plotted for the whole range of $Pe$, while analytical solutions from regular perturbation (small $Pe$) have only been plotted for small values of $Pe$, upto 0.8. Again reasonably good agreement between the two is observed for low values of $Pe$. The inset depicts the magnified variation in $\bar{u}_\infty$ with $Pe$ from 2.5 to 4.8, for $\beta = 2$. The other relevant parameters are given by: $\zeta_0 = 0.3$, $n = 2$ and $\bar{\delta} = 0.5$. (b) Plot of $\bar{u}_\infty$ vs $Pe$, for three different values of $\beta$, i.e., the field strength, 0.3, 0.5 and 1.0. The other parameters are given by: $\zeta_0 = 0.03$, $n = 2$ and $\delta = 0.03$. Numerical solutions along with the analytical solutions from the matched asymptotics ($Pe \sim O(1)$ and small $\delta$) have been plotted, for comparison. We note that they agree well, especially for lower values of $\beta$.

### 7.2.2. Effects of relaxing the "weak field" approximation

Figures 5.(a) and (b) demonstrate the variation in the free stream velocity, with Peclet number, for different levels of applied axial field strength, while the values of all other relevant parameters have been mentioned in the caption. In figure 5.(a), the analytical solutions based on the weak advection limit (section 4) have been plotted, while in (b),

approximate solutions from thin EDL limit have been shown for comparison. In the inset of figure 5.(a), we show the magnified variation in the free stream velocity for $\beta = 2$ and $Pe = 2.5 - 4.8$. These figures embody the effects of both ionic advection and strong axial field on the flow dynamics of the system.

We first concentrate on figure 5.(a) and note several interesting points. The first thing to observe is that reasonably good agreement is observed between the analytical and numerical solutions till $Pe \sim 0.5$. Second, the decrement in $\bar{u}_\infty$ is larger for larger values of $\beta$. This can be explained easily from the asymptotic solutions, in section 4. We recall that, the effect of $\beta$ comes into the flow through the $O(Pe)$ and $O(Pe^2)$ equations of fluid flow (Refer to eqns. 4.8 – 4.10 and 4.17 – 4.19). Therefore, an increase in $\beta$ would naturally lead to increased contributions from the higher order velocities. Since the higher order velocities tend to decrease the free stream velocity, increase in $\beta$ subsequently leads to a decrease in the same. The variation in $\bar{u}_\infty$ with $Pe$ is, however, more interesting. We note that for lower values of $Pe$, the free stream velocity decreases rapidly, reaches a minima for $Pe \sim O(1)$ and then increases very slowly as $Pe$ is increased. The same trend is observed for both values of $\beta$, as evident from the inset. Such an unusual behavior is difficult to explain, based on the solutions of weak advection regime. However, we note that a very qualitative and rough explanation can indeed be presented based on the solutions in the thin EDL limit, since the maxima in the free stream velocity is reached for $Pe \sim O(1)$ values. We note from (5.54) that the free stream velocity in the thin EDL limit, shows a quadratic variation in $Pe$ (till $O(\delta^2)$), where the coefficient of $Pe^2$ is negative. Hence, the variation of $\bar{u}_\infty$ with $Pe$ would behave like an upside down parabola, with a peak value. Loosely speaking, this is the trend, which we observe in figure 5.(a). From a physical point of view, we observe that, for a given potential, there is only a finite amount of charge in the fluid and thus only a limited amount of change in the charge distribution can be induced through the fluid velocities. Therefore, the free stream velocities cannot be reduced indefinitely by increasing the Peclet number.

Figure 5.(b) showcases very similar variations in $\bar{u}_\infty$, as in the previous figure. We note that the analytical and numerical solutions agree reasonably well, while the agreement is better for lower values of $\beta$. Note that, for $\beta = 1$, $n = 2$ and $Pe \sim 3$, the $O(\delta)$ contribution to $\bar{u}_\infty$ remains $O(1)$. But, the $O(\delta^2)$ contribution becomes $\sim 30$, because of the presence of terms like $Pe^2$ and $n^3$, thus exceeding $O(1)$ magnitude by a large extent. This is expected from a divergent singular perturbation expansion, as the one presented in (5.54). Therefore, we cannot expect the analytical solutions to show good agreement with the numerical solutions, under these relatively extreme scenarios. Another interesting point to note from the present figure is that the free stream velocity shows monotonic behavior with $Pe$ and the minima in the velocity, as observed for 5.(a), is absent here. Our numerical simulations suggest that such extremas do appear in the thin EDL limit, although for significantly higher values of $Pe$. Therefore, we have not shown those data points in the present figure.

Figures 6.(a) and (b) explicitly explore the effect of applied axial field strength on the free stream velocity, for a number of Peclet number values. To this end, we first concentrate on figure 6.(a), where, variations in $\bar{u}_\infty$ as a function of $\beta$ for two different values of $Pe = 0.3$

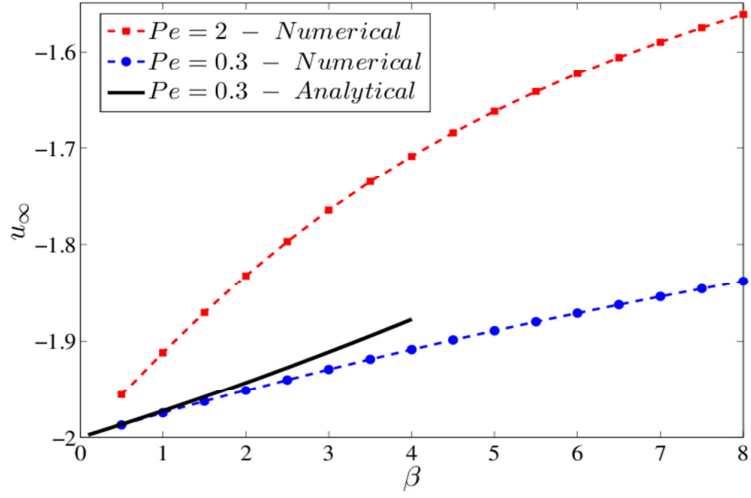

(a)

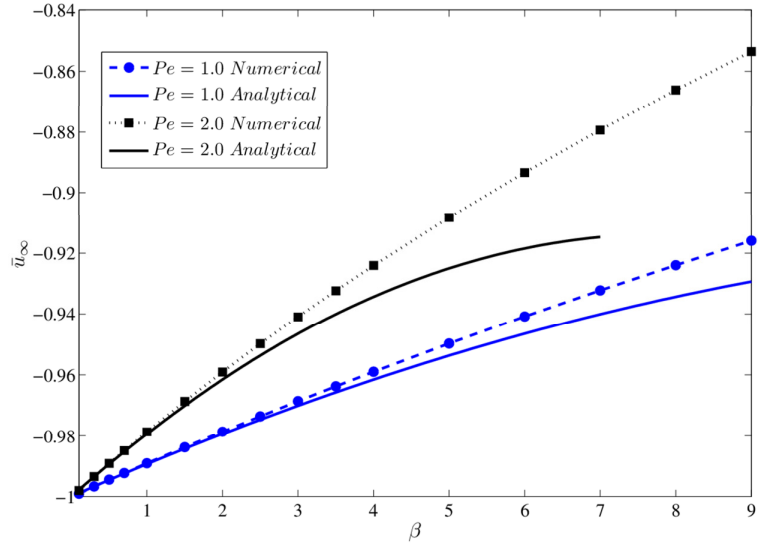

(b)

Figure 6: (a) $\bar{u}_\infty$ vs $\beta$, for two different values of $Pe$ = 0.3 and 2. Analytical solutions (based on regular perturbation for small $Pe$, section 4) have been plotted for $Pe$ = 0.3, upto $\beta$ = 4. Again reasonably good agreement between the two is observed for values of $\beta$ close to 1. The other relevant parameters are given by: $\zeta_0$ = 0.1, $n$ = 2 and $\bar{\delta}$ =0.5. (b) Plot of $\bar{u}_\infty$ vs $\beta$ (the field strength), for two different values of $Pe$ = 1 and 2 in the thin EDL regime. The values of the other parameters are given by: $\zeta_0$ = 0.1, $n$ = 1 and $\delta$ = 0.025. Numerical solutions along with the analytical solutions from the matched asymptotics ($Pe \sim O(1)$ and small $\delta$) have been plotted, for comparison. We note that they show reasonably good agreement, even for larger values of $\beta$.

and 2 have been depicted, while the values of the other relevant parameters have been mentioned in the caption. For $Pe$ = 0.3, we have plotted the analytical solutions as obtained from the weak advection regime (section 4). This figure is in agreement with the previous one, since it shows that the free stream velocity decreases as $\beta$ is increased. In other words, decreasing the external field strength diminishes the free stream velocity further. This trend, as predicted by the numerical solutions is again in complete agreement with that predicted by the analytical solutions. Reasonably good quantitative agreement is also observed between the two solutions, till $\beta \sim 2.5$. From the analytical solutions, it is very clear that the effect of $\beta$ explicitly comes into play through $O(Pe)$ and $O(Pe^2)$ contributions, in the weak advection regime. We have already demonstrated that these higher order terms increase the periodicity of the flow at the expense of free stream velocity. Therefore, increased values of $\beta$ would naturally lead to decreased free stream velocity, increasing recirculation in the process. Figure 6.(b) investigates a similar variation in $\bar{u}_\infty$ with $\beta$, albeit in the thin EDL regime, for two different values of $Pe = 1$ and 2, while the values of the other relevant parameters have been mentioned in the caption. We have plotted the approximate solutions in the thin EDL regime (section 5) for comparison. Reasonably good agreement is observed between the numerical and analytical solutions, with the agreement being better for $Pe = 1$. We note that, this figure depicts very similar qualitative trends to the previous one, wherein, increment in the axial field strength (or, a decrement in $\beta$) causes the free stream velocity to go up. We reiterate that, such behavior of the free stream velocity can be easily predicted from equation (5.54), wherein the effects of field strength and advection strength are seen to play opposite roles, in determining the free stream velocity, in the first order of $\delta$.

Combining figures 5 and 6, we infer that a strong axial field increases the "*free stream velocity*", whereas, larger advection strength decreases the same. One can thus conclude that, applied field strength and ionic advection have opposite effects on the flow dynamics, when varied separately, keeping the other one constant.

### 7.3. Flow Dynamics

We begin our discussions with figure 7, where the streamline patterns obtained from numerical (shown in (a.1) and (b.1)) and analytical solutions (shown in (a.2) and (b.2)) are compared, while the values of the other relevant parameters have been mentioned in the caption. In figures (a), we compare the analytically obtained streamline structures from the thin EDL limit with the numerical solutions ($Pe = 5$), while in figure (b), the analytical solution in the weak advection regime ($Pe = 0.2$) have been compared. Good agreement is observed between the numerical and analytical solutions in both the cases. The streamline patterns are quite standard for electroosmotic flows in presence of modulated charges (Ghosh & Chakraborty 2012; Ajdari 1995). For thicker EDLs (figures (b)), a single recirculation roll near the wall is observed, while the streamlines away from the wall have less curvature indicating free stream velocity. However, for thinner EDLs (figures (a)), a smaller secondary roll can be seen above the edge of the EDL. The streamlines for the case of thin EDL, with $Pe = 5$, are more heavily distorted as compared to the thick EDL-weak advection regime. This can be attributed to the fact that, inclusion of advection adds more periodicity to the flow, at the expense of net fluid movement. These higher order periodic terms, hence cause the streamlines to get distorted, the distortion being stronger as we increase the advection strength. Therefore, we naturally expect the streamlines to show higher levels of undulations for the cases of thinner EDLs, with $Pe = 5$. It is important to note here that, the present figure depicts that ionic advection indeed has negligible effects on the streamline patterns, even for $Pe \sim O(0.1)$ (classified as, thick EDLs and weak advection limit). Therefore, for similar or smaller $Pe$ values, one can safely neglect the effect of ionic advection and start with the Boltzmann distributions. However, we subsequently demonstrate that for larger values of $Pe$, one can no longer overlook the effects of ionic advection, even for relatively thick EDLs and has to start from a more primitive PNP model.

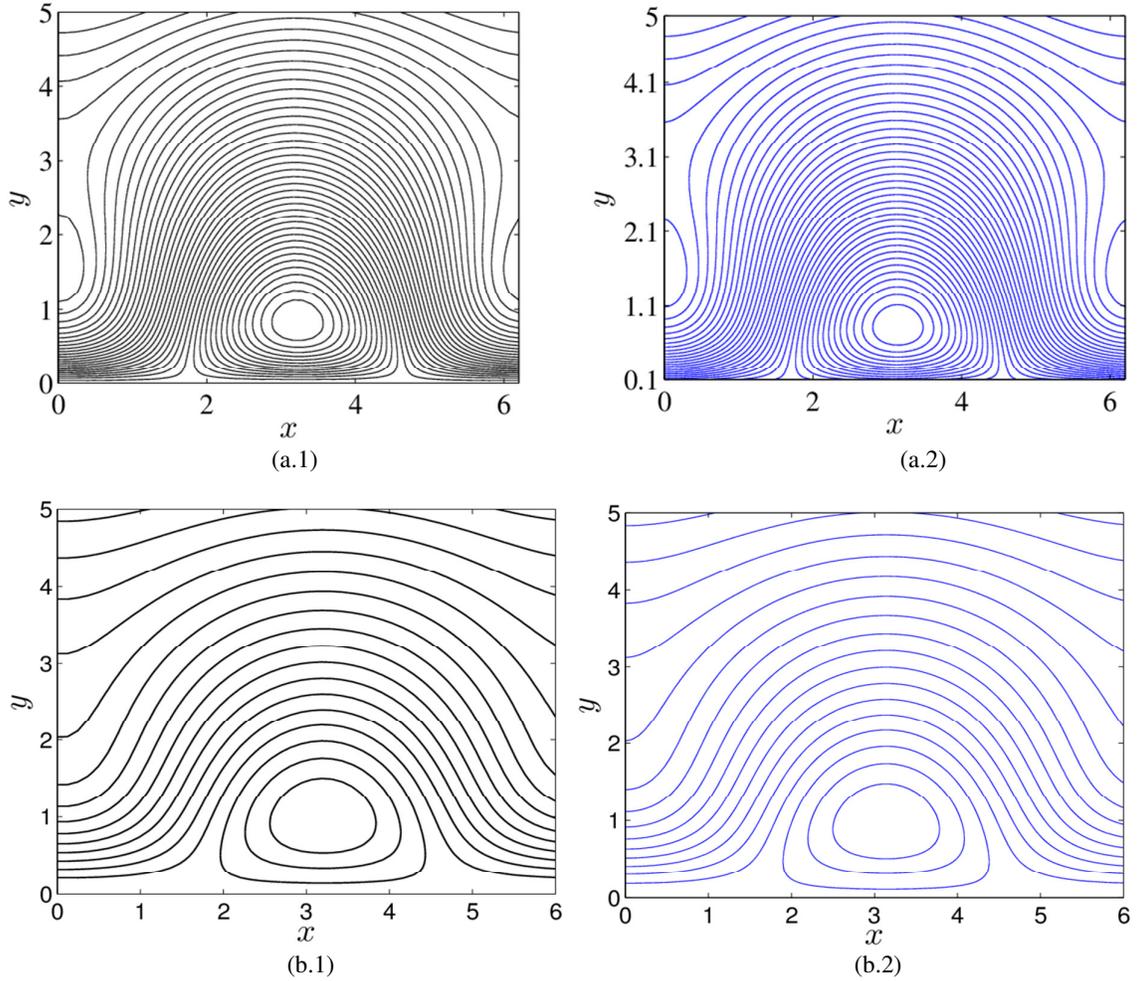

Figure 7: (a) Comparison between streamlines as obtained from (a.1) numerical and (a.2) analytical solutions (as obtained from matched asymptotics), for a representative case of, $Pe = 5$, $n = 0.1$, $\zeta = 0.1$, $\delta = 0.05$ and $\beta = 1$. The agreement between the two streamline structures is very good. Note that, in figure (a.2), we have only shown the streamlines from the solutions in the outer region (i.e., the bulk). (b) Streamlines depicting the flow pattern, for $\beta = 2$, $Pe = 0.2$, $\bar{\zeta}_0 = 0.1$, $n = 0.1$ and $\delta = 0.3$. Figure (b.1) shows the numerical solutions, while (b.2) demonstrates the analytical solution as obtained from regular perturbation analysis (for small $Pe$, section 4). Good agreement is observed between the two solutions.

### *7.3.1. Effects of Ionic advection*

Figures 8.(a) – (d) explore the effects of advection strength on streamline patterns (numerical only), while the other relevant parameters have been mentioned in the caption. An obvious point to note from this figure is that increasing the Peclet number increases the size of the roll residing near the wall. We further note that the streamlines just around the recirculation roll get distorted to large extents as $Pe$ is augmented. For $Pe = 10$, we note that secondary recirculation rolls appear slightly outside the EDL (at $y \sim 2$). The trend described above can be very easily explained qualitatively based on the asymptotic solutions. These solutions show that, inclusion of advection increases the periodicity in the flow at the expense

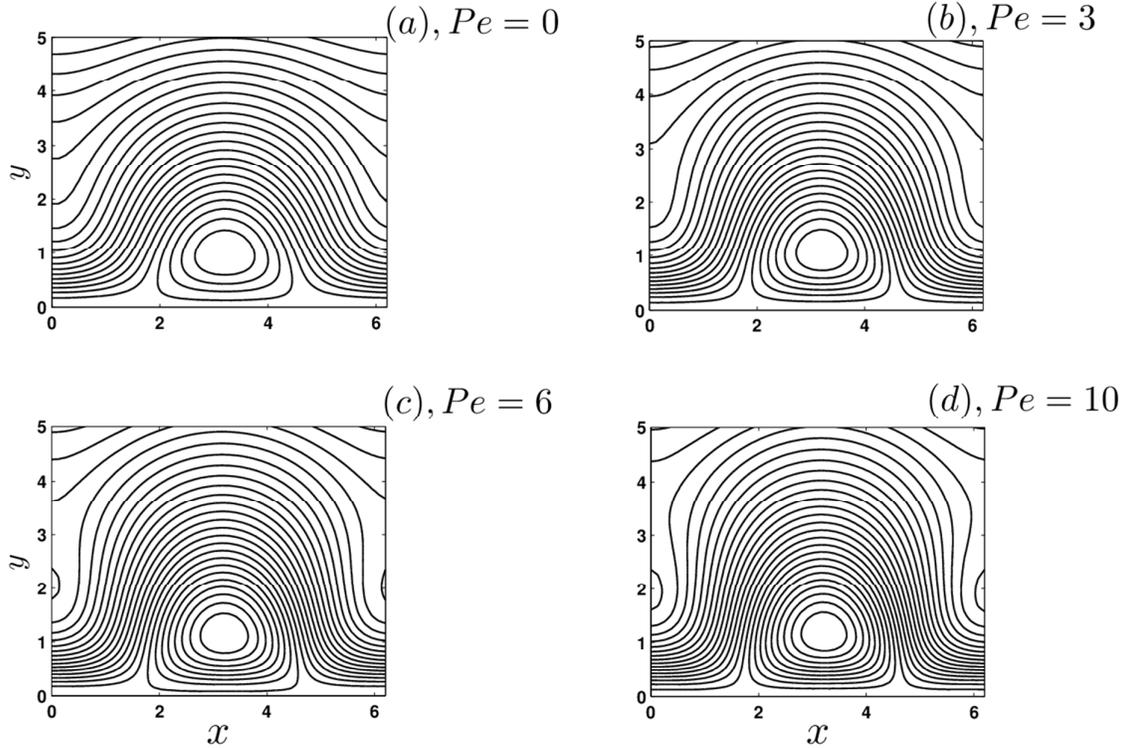

Figure 8: Streamlines (numerical only) depicting the flow patterns for different Peclet numbers: (a) $Pe = 0$, (b) $Pe = 3$, (c) $Pe = 6$ and (d) $Pe = 10$. The rest of the parameters have the following values: $\beta = 4$, $\bar{\zeta}_0 = 1$, $n = 0.1$ and $\delta = 0.3$.

of "free stream velocity", which finally results in distorted streamlines and larger recirculation rolls.

### 7.3.2. Effects of relaxing the "weak field" approximation

We next move to figures 9.(a) – (d), where the effect of applied external field strength on the flow dynamics has been investigated, in presence of ionic advection. To this end, we have shown the flow pattern (numerical only) for four different values of $\beta$, while the values of the other relevant parameters have been mentioned in the caption. The present figures again demonstrate similar qualitative behavior as depicted in figures 8.(a) – (d). Increasing $\beta$ has the same qualitative effect as increasing $Pe$, which include larger recirculation rolls, more distorted streamlines near the walls and appearance of secondary rolls above the EDL. We again take clue from the asymptotic solutions and note that effect of $\beta$ in the equation of motion appears in $O(Pe)$ and $O(Pe^2)$ through the potentials at the same order (i.e., $\varphi_1$ and $\varphi_2$) in the weak advection regime. On the other hand, $\beta$ also influences the solutions analogously in the thin EDL limit, through the $O(\delta)$ and $O(\delta^2)$ corrections. Therefore, we expect $\beta$ to have similar qualitative effects as $Pe$, at least for low to moderate values of $Pe$. In that sense, this figure is in agreement with figure 6, which indicates that $\bar{u}_\infty$ decreases as $\beta$ is increased. We naturally draw the conclusion that a decreased free stream velocity should be accompanied by higher periodicity in the flow, as noted in the present figure. Therefore, based on figures 6 and 9, we can state that decreasing the strength of the externally applied axial potential (indicated by an increase in $\beta$) leads to increased periodicity in the flow, at the expense of free stream velocity (or, net throughput in case of a confinement). This statement in fact, agrees with the previously drawn conclusion that, applied field strength and advection strength tend to have opposite effects on the flow dynamics.

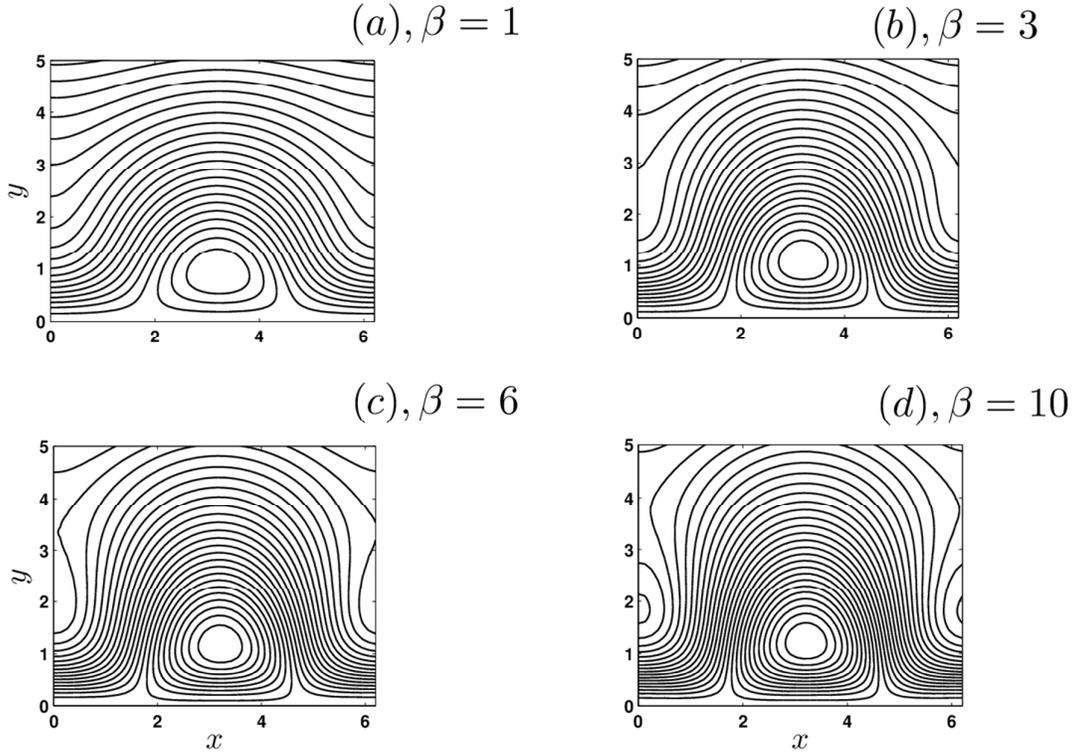

Figure 9: Stream line patterns (numerical only) depicting the flow dynamics, for four different values of $\beta$: (a) 1, (b) 3, (c) 6 and (d) 10. The other relevant parameters have the following values: $\bar{\zeta}_0$ =1, $Pe$ = 5, $n$ = 0.1 and $\delta$ = 0.3.

## 8. Conclusions

In the present study, we have analyzed the effect of ionic advection on the flow dynamics, potential distribution and related electrokinetic phenomena, in presence of modulated zeta potential on the surface, beyond weak field limits. We have performed numerical simulations, for a wide range of ionic Péclet number and relative strength of the axial electric field, beyond the Debye-Hückel limit for the surface potential. We have further analyzed two separate asymptotic limits, namely, the weak advection limit, where we assume $Pe \ll 1$, while the EDL thickness can be arbitrary, without exceeding O(1) in magnitude. The second asymptotic limit addressed in the present study is the thin EDL limit, with $\delta \ll 1$ and $Pe \sim O(1)$ or less. We have demonstrated that the numerical and analytical solutions show reasonably good agreement within proper parametric range.

There are several interesting conclusions to be drawn from the present analysis. First and foremost, we have depicted that inclusion of advection actually slows down the net flow, increasing the periodicity in the process. Similarly, decreasing the relative strength of the external electric field (or, in other words increasing the parameter $\beta$ beyond the limiting small values, typically used under weak field approximation) has almost identical effects on the flow dynamics. We therefore, concluded that ionic advection strength and external field strength have opposite effects on the overall flow dynamics. A second conclusion to be drawn from the present analysis is that the EDL thickness strongly influences the free stream velocity, when advection is taken into account. The maximum change in the free stream velocity occurs for a particular EDL thickness (here, for $\delta \sim 0.3$) for almost all values of $Pe$. This essentially indicates that one cannot use a simple Smoluchowski like slip velocity condition to model electroosmotic flows in presence of thin EDLs, in sharp contrast to what is classically done when ionic advection is neglected altogether. A final point to note is that, in absence of ionic advection, the potential ($\varphi$) diminishes quickly outside the EDL, irrespective of the flow. However, when charge advection is accounted for, the potential drops at a slower rate, thus assuming a non-zero value even beyond the EDL. We have demonstrated that such variations in the potential can be easily explained from our thin EDL solutions, wherein an overall charge balance criteria requires an induced potential, outside the EDL. We have further shown that this potential has dominant contribution at $O(\delta)$, for thin EDLs. In addition to this, we have also shown that in the weak advection regime, i.e., for $Pe \sim O(0.1)$ or less, the effects of ionic advection on the overall flow dynamics is in fact negligible (although it influences the free stream velocity). Therefore, for $Pe$ values in the aforementioned range, one can safely use the Boltzmann distribution for describing the charge density in the flow. For higher values of $Pe$, however, one has to take into account the effects of advection, even for relatively thicker EDLs ($\delta > 0.1$), in order to obtain an accurate description of the flow field.

**Appendix – A: Expressions for functions in section 4**

Expressions for $Y_1^{(1)}$ and $Y_2^{(1)}$ in section 4:

$$Y_1^{(1)} = \frac{2\bar{\zeta}_0 n \left[ \left\{ \delta_1 (\delta_1 - \delta) e^{-\frac{\bar{y}}{\delta_1}} + ((1+\bar{y})\delta_1 - \bar{y}) e^{-\bar{y}} \delta \right\} e^{-\frac{\bar{y}}{\delta}} - e^{-\frac{\bar{y}}{\delta_1}} \delta_1^2 \right]}{\delta_1^2} \qquad (A.1)$$

$$Y_2^{(1)} = -\frac{\bar{\zeta}_0 \delta^2 \bar{y} (\delta_1 - 1)^2 e^{-\bar{y}(1+1/\delta_1)}}{\delta_1^3} \qquad (A.2)$$

Expressions for $U_k^{(c)}, U_k^{(s)}$ & $V_k^{(s)}$ $(k = 1, 2...etc)$ in section (4.2):

$$U_1^{(c)} = \frac{\beta e^{-\frac{\bar{y}}{\delta_1}} \delta_1^2 \delta^2 \left( 2g_2(\bar{y}) - \frac{d^2 g_2}{d\bar{y}^2} \right) + 2\beta n \delta_1^2 g_1(\bar{y}) e^{-\frac{\bar{y}}{\delta}} + 2\beta \delta^2 g_2(\bar{y}) e^{-\frac{\bar{y}}{\delta_1}}}{2\delta^2 \delta_1^2} \qquad (A.3)$$

$$U_2^{(c)} = \frac{\beta e^{-\frac{\bar{y}}{\delta_1}} \delta_1^2 \delta^2 \left( -2g_1(\bar{y}) + \frac{d^2 g_1}{d\bar{y}^2} \right) + 4\beta n \delta_1^2 g_2(\bar{y}) e^{-\frac{\bar{y}}{\delta}} + 2\beta \delta^2 g_1(\bar{y}) e^{-\frac{\bar{y}}{\delta_1}}}{2\delta^2 \delta_1^2} \qquad (A.4)$$

$$U_3^{(c)} = \frac{\beta e^{-\frac{\bar{y}}{\delta_1}} \delta_1^2 \delta^2 \left(-6g_2(\bar{y}) + \frac{d^2 g_2}{d\bar{y}^2}\right) + 2\beta \delta^2 g_2(\bar{y}) e^{-\frac{\bar{y}}{\delta_1}}}{2\delta^2 \delta_1^2}, \quad U_k^{(s)} = k^2 g_k(\bar{y}) - \frac{d^2 g_k}{d\bar{y}^2}; \ k = 1, 2$$

(A.5)

$$V_1^{(s)} = -\frac{\beta}{2\delta^2 \delta_1^2} \left[ \begin{array}{l} 2\delta \delta_1^2 n e^{-\frac{\bar{y}}{\delta}} \left(\frac{d^2 g_1}{d\bar{y}^2} - g_1\right) + \delta \delta_1^2 e^{-\frac{\bar{y}}{\delta_1}} \left(\frac{d^2 g_2}{d\bar{y}^2} - g_2\right) + \\ e^{-\frac{\bar{y}}{\delta_1}} \delta^2 \frac{dg_2}{d\bar{y}} (\delta_1 - 1)^2 - 2n\delta_1^2 e^{-\frac{\bar{y}}{\delta}} \frac{dg_1}{d\bar{y}} \end{array} \right]$$

(A.6)

$$V_2^{(s)} = -\frac{\beta}{2\delta^2 \delta_1^2} \left[ \begin{array}{l} 2\delta \delta_1^2 n e^{-\frac{\bar{y}}{\delta}} \left(-\frac{d^2 g_1}{d\bar{y}^2} + g_1\right) + 2\delta \delta_1^2 e^{-\frac{\bar{y}}{\delta_1}} \left(4g_2 - \frac{d^2 g_2}{d\bar{y}^2}\right) + \\ e^{-\frac{\bar{y}}{\delta_1}} \delta^2 \frac{dg_1}{d\bar{y}} (1 - \delta_1^2) + 2n\delta_1^2 e^{-\frac{\bar{y}}{\delta}} \frac{dg_1}{d\bar{y}} \end{array} \right]$$

(A.7)

$$V_3^{(s)} = \frac{\beta e^{-\frac{\bar{y}}{\delta_1}}}{2\delta^2 \delta_1^2} \left[ 2\delta \delta_1^2 \left(\frac{d^2 g_2}{d\bar{y}^2} + g_2\right) + \delta^2 \frac{dg_2}{d\bar{y}} (1 - \delta_1^2) \right]$$

(A.8)

**Appendix – B: Derivation of overall charge balance condition**

Here, we derive the condition (5.12), used to balance the overall charge density inside the EDL. To this end, we note that, there are two separate overall balance equations, one for the total charge distribution ($\rho_e$) and the other for the total ionic concentration ($c$). However, since these conditions are used solely to derive analytical expressions in the present study, we only derive the overall balance condition for charge density $\rho_e$, on account of the fact that for analytical solutions we have assumed $c = 2$. We shall apply the very fundamental Reynolds' Transport Theorem (RTT) (Kundu & Cohen 2004) to the conservation of charges, inside the EDL. For this purpose, we first identify a suitable control volume (CV) and the corresponding fluxes in and out of the control volume. The choice of the CV has been depicted in figure 10. It is simple rectangular CV, of width $dx$ (in the limit, $dx \to 0$), with its edges lying parallel to the $x$ and $y$ axes. The upper edge of the CV lies at the edge of the EDL, while the bottom edge lies on the wall. The statement of RTT, for charge density is as follows (in absence of any charge generation):

$$\begin{bmatrix} \textit{Rate of charge} \\ \textit{in the total charge density} \\ \textit{inside the CV} \end{bmatrix} = \begin{bmatrix} \textit{Rate of charge} \\ \textit{inflow to the CV} \end{bmatrix} - \begin{bmatrix} \textit{Rate of charge} \\ \textit{outflow form the CV} \end{bmatrix} \quad (B.1)$$

If the overall charge density per unit length in $x$ direction is denoted by $q$, then we have, $q = \int_{EDL} ze(c_1 - c_2) dy$. We now consider the fluxes along various faces. To this end, we note that the flux across the face AB is influx along $x$ direction. The flux density is given by: $J_x|_{AB} = zec_0 \left( u\bar{\rho}_e - D \frac{\partial \bar{\rho}_e}{\partial x} - \frac{zeD}{kT} \bar{c} \frac{\partial \psi}{\partial x} \right)$, with $\psi$ being the total potential (see section 2). Similarly the flux density across the face CD, which is a net outflux, is given by:

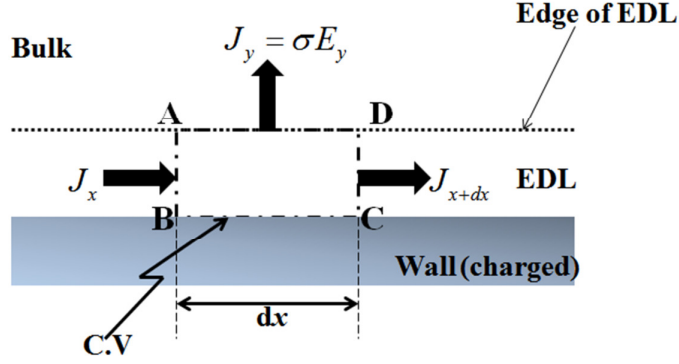

Figure 10: Schematic for the derivation of overall charge balance condition in Appendix-C. Here, ABCD represents the faces of the C.V., which has a width d$x$, along the $x$ direction. The top face, DA is coincident with the edge of the EDL, while the bottom surface (BC) lies on the wall. The other two faces, i.e., AB and CD are vertically oriented. The fluxes and their directions across the faces have been shown in the schematic.

$J_x|_{CD} = J_x|_{AB} + \dfrac{\partial J_x|_{AB}}{\partial x} dx$. There is no net flux along the face BC. Finally, the flux across the face DA is in the $y$ direction and is caused by a current driven by the outer layer potential. Since this face is the edge of the EDL, the current density is simply given by the expression, $J_y|_{DA} = \sigma E_y|_{\bar{y}\to 0}$ and it acts as an outflow of ions. In the foregoing expression, $\sigma$ is the conductivity of the bulk, or, the outer solution and $E_y$ is the $y$-component of the electric field in the bulk, at the edge of the EDL ($\bar{y}\to 0+$). We now accumulate all the fluxes in (B.1) to write the conservation equation for the overall charge density. This equation reads:

$$\dfrac{\partial q}{\partial t} dx = -\int_{EDL} \dfrac{\partial J_x|_{AB}}{\partial x} dx dy - \sigma E_y|_{\bar{y}\to 0} dx \qquad (B.2)$$

Plugging in the expressions for fluxes, we get (note that $\psi = \varphi - E_0 x$),

$$\dfrac{\partial q}{\partial t} = -zec_0 \int_{EDL} \dfrac{\partial}{\partial x}\left(u\bar{\rho}_e - D\dfrac{\partial \bar{\rho}_e}{\partial x} - \dfrac{zeD}{kT}\bar{c}\left(\dfrac{\partial \varphi}{\partial x} - E_0\right)\right) dy - \lim_{\bar{y}\to 0}\sigma E_y \qquad (B.3)$$

We now non-dimensionalize (B.3) with appropriate scales. To this end, we choose, $\tilde{q} = q/q_c$, $\bar{t} = t/t_c$; $U = u/u_{HS}$, $X = x/d$; $\bar{E}_y = E_y/E_c$; $\sigma = 2c_0 z^2 e^2 D/(kT)$, $Y = y/\lambda_D$ and $\tilde{\varphi}, \bar{\varphi} = \varphi/\bar{\zeta}_0$. Here, we have taken, $q_c = 2zec_0\bar{\zeta}_0\lambda_D$, $E_c = \bar{\zeta}_0/d$, $t_c = \lambda_D d/D$, where $\lambda_D$ is the characteristic EDL thickness. The other relevant scales have already been discussed in section 2. Enforcing the foregoing non-dimensionalization scheme in (B.3), we deduce:

$$\dfrac{\partial \tilde{q}}{\partial \bar{t}} = -\dfrac{\delta}{2\bar{\zeta}_0}\int_0^\infty \dfrac{\partial}{\partial X}\left(PeU\tilde{\rho}_e - \dfrac{\partial \tilde{\rho}_e}{\partial X} - \bar{\zeta}_0 \bar{c}\left(\dfrac{\partial \tilde{\varphi}}{\partial X} - \dfrac{1}{\beta}\right)\right) dY + \lim_{\bar{y}\to 0}\dfrac{\partial \bar{\varphi}}{\partial \bar{y}} \qquad (B.4)$$

Note that, we have used "tilde" sign for representing the variables inside the EDL. The limit of integration changes from 0 to infinity, since in terms of rescaled variable $Y$, the edge of the EDL is approached as, $\lim Y \to \infty$. Additionally, we have used the RC time scale to non-dimensionalize time (Bazant et al. 2004). This is on account of the fact that, for systems showing temporal variation in the EDL structure, the RC time scale remains the natural response time scale of the system (Bazant et al. 2004; Højgaard Olesen et al. 2010). We reiterate that, in (B.4), $\bar{\zeta}_0 = ze\zeta_0/kT$ and $\tilde{q} = \int_0^\infty (\bar{\rho}_e/2\bar{\zeta}_0)dY$. Equation (B.4) is very similar to the overall charge balance equation used previously by Schnitzer and coworkers (Yariv et al. 2011; Schnitzer et al. 2012), Ramos et al (Ramos et al. 2003) and Storey et al. (Storey et al. 2008). We further note that, in absence of charge advection, assuming the EDL remains in quasi-equilibrium, (B.4) reduces to the form previously employed by Ajdari (Ajdari 2000) and Storey et al. (Storey et al. 2008). We now enforce the approximations used in the present study, namely, $\bar{c} \approx 2$, as attributable to low magnitude of surface potential. Finally replacing $\tilde{\rho}_e$ with $2\bar{\zeta}_0\tilde{\chi}$, one can write for steady state motion ($\partial \tilde{q}/\partial \bar{t} = 0$):

$$\lim_{\bar{y}\to 0}\frac{\partial \bar{\varphi}}{\partial \bar{y}} = \delta\int_0^\infty \frac{\partial}{\partial X}\left(PeU\tilde{\chi} - \frac{\partial \tilde{\chi}}{\partial X} - \frac{\partial \tilde{\varphi}}{\partial X}\right)dY \tag{B.5}.$$

This is exactly same as equation (5.12).

**Appendix – C: Expressions for $U_2$ and $V_1$ in section 5**

The Y-velocity $V_1$ at O($\delta$) can be easily derived by integrating the continuity equation, subject to no penetration boundary condition at the wall. The resulting velocity has the following expression:

$$V_1 = \sum_{k=1}^{3}\widehat{V}_s^{(k)}(Y)\sin(kX) + \sum_{k=1}^{2}\widehat{V}_c^{(k)}(Y)\cos(kX) \tag{C.1}.$$

In (C.1), the various functions have the following expressions:

$$\widehat{V}_s^{(1)} = Pe\beta\left(n^2 + \frac{1}{4}\right)\left(e^{-Y} + Y - 1\right) + Y^2; \quad \widehat{V}_s^{(2)} = Pe\beta n\left(2Y + e^{-Y} - 1\right) \tag{C.2}$$

$$\widehat{V}_s^{(3)} = \frac{3}{4}Pe\beta\left(Y + e^{-Y} - 1\right); \quad \widehat{V}_c^{(1)} = nPe\left(1 - Y - e^{-Y}\right); \quad \widehat{V}_c^{(2)} = \frac{Pe}{2}\left(1 - Y - e^{-Y}\right) \tag{C.3}.$$

Expressions for $W_c^{(k)}\text{'s}$ and $W_s^{(k)}\text{'s}$ in section 5.5:

$$W_n^{(2)} = \frac{1}{8}\beta e^{-Y}\left(3Pene^{-Y} + 4nPeY + 8nPe - 4g_s^{(1)} + nPeYe^{-Y}\right) \tag{C.4}$$

$$W_c^{(1)} = \begin{bmatrix} n^2Pe\beta\left(\dfrac{7}{16}e^{-2Y} + \dfrac{1}{8}Ye^{-2Y} + 3e^{-Y} + Ye^{-Y}\right) - n\beta g_s^{(1)}e^{-Y} \\ + e^{-Y}\beta\left(-g_s^{(2)} + \dfrac{1}{2}PeY + \dfrac{3}{32}PeYe^{-Y} + \dfrac{19}{64}Pee^{-Y} + \dfrac{5}{4}Pe - \dfrac{Y}{2\beta}\right) + e^{-Y}\left(1 - g_c^{(1)}\right) \end{bmatrix} \tag{C.5}$$

$$W_c^{(2)} = \begin{bmatrix} \beta e^{-Y}\left(-\dfrac{1}{2}g_s^{(1)} - \dfrac{3}{2}g_s^{(3)} - 2ng_s^{(2)} + \dfrac{1}{2}nPee^{-Y} + 5nPe + \dfrac{3}{2}nPeY + \dfrac{1}{8}nPee^{-Y}\right) \\ -g_c^{(2)}e^{-Y} \end{bmatrix} \tag{C.6}$$

$$W_c^{(3)} = \begin{bmatrix} Pe\beta e^{-Y}\left(\frac{1}{2}Y + \frac{7}{4} + \frac{9}{64}e^{-Y} + \frac{1}{32}Ye^{-Y}\right) \\ -\beta e^{-Y}\left(2g_s^{(4)} + 3ng_s^{(3)} + g_s^{(2)}\right) - g_c^{(3)}e^{-Y} \end{bmatrix} \quad \text{(C.7)}$$

$$W_c^{(4)} = -\frac{1}{2}\beta e^{-Y}\left(3g_s^{(3)} + 8ng_s^{(4)}\right); \quad W_c^{(5)} = -2\beta g_s^{(4)}e^{-Y} \quad \text{(C.8)}$$

$$W_s^{(1)} = \left[-\frac{1}{8}\beta Pe^2 n e^{-Y} - \frac{1}{8}Pee^{-Y}\left(10nY + 2nY^2\right) + \beta e^{-Y}\left(g_c^{(2)} + ng_c^{(1)}\right) - g_s^{(1)}e^{-Y}\right] \quad \text{(C.9)}$$

$$W_s^{(2)} = \left[\frac{1}{2}\beta n^2 Pe^2 e^{-Y} - \frac{1}{8}Pee^{-Y}\left(5Y + Y^2\right) + \frac{e^{-Y}}{2}\left\{\beta\left(4ng_c^{(2)} + 3g_c^{(3)} + g_c^{(1)}\right) - 2g_s^{(2)}\right\}\right] \quad \text{(C.10)}$$

$$W_s^{(3)} = e^{-Y}\left[\frac{\beta}{8}\left(8g_c^{(2)} + 24ng_c^{(3)} + 3nPe^2\right) - g_s^{(3)}\right]; \quad W_s^{(4)} = \frac{1}{16}e^{-Y}\left(Pe^2\beta + 24\beta g_c^{(3)} - 16g_s^{(4)}\right) \quad \text{(C.11)}$$

Expressions for the constants in $U_2$ are given by:

$$c_{2,2}^{(1)} = -1 + g_c^{(1)} + \beta\left(-\frac{99}{64}Pe + ng_s^{(1)} + g_s^{(2)} - \frac{55}{16}n^2 Pe\right) \quad \text{(C.12)}$$

$$c_{2,2}^{(2)} = g_c^{(2)} + \beta\left(\frac{1}{2}g_s^{(1)} + \frac{3}{2}g_s^{(3)} + 2ng_s^{(2)} - \frac{11}{2}nPe\right) \quad \text{(C.13)}$$

$$c_{2,2}^{(3)} = g_c^{(3)} + \beta\left(2g_s^{(4)} - \frac{121}{64}Pe + 3ng_s^{(3)} + g_s^{(2)}\right); \quad c_{2,2}^{(4)} = \frac{1}{2}\beta\left(3g_s^{(3)} + 8ng_s^{(4)}\right); \quad c_{2,2}^{(5)} = 2\beta g_s^{(4)} \quad \text{(C.14)}$$

$$s_{2,2}^{(1)} = g_s^{(1)} + \beta\left(\frac{1}{8}n^2 Pe - ng_c^{(1)} - g_c^{(2)}\right); \quad \text{(C.15)}$$

$$s_{2,2}^{(2)} = g_s^{(2)} - \beta\left(\frac{1}{2}n^2 Pe^2 + 2ng_c^{(2)} + \frac{1}{2}g_c^{(1)} + \frac{3}{2}g_c^{(3)}\right) \quad \text{(C.16)}$$

$$s_{2,2}^{(3)} = g_s^{(3)} - \beta\left(\frac{3}{8}Pe^2 n + 3ng_c^{(3)} + g_c^{(2)}\right); \quad s_{2,2}^{(4)} = g_s^{(4)} - \beta\left(\frac{1}{16}Pe^2 + \frac{3}{2}g_c^{(3)}\right) \quad \text{(C.17)}$$